\documentclass[aps,showpacs,prb,reprint,longbibliography,superscriptaddress]{revtex4-2}

\usepackage[colorlinks=true,linkcolor=blue,citecolor=blue,urlcolor=blue]{hyperref}
\usepackage{setspace} 
\usepackage{graphicx}
\usepackage{amsmath}
\usepackage{color}
\usepackage{amsmath}
\usepackage{mathtools}
\usepackage{amssymb}
\usepackage{verbatim}
\usepackage{physics}
\usepackage{latexsym}
\usepackage{enumerate} 
\usepackage{bm} 
\usepackage[caption=false]{subfig}
\usepackage{multirow}
\usepackage{booktabs}
\usepackage{siunitx}

\newcommand*{\Ang}{\, \mbox{\normalfont\AA}}

\usepackage{float}

\usepackage{lipsum}

\begin{document}

\title{\textbf{\Large{Phonon screening and dissociation of excitons at finite temperatures from first principles}}}

\author{Antonios M. Alvertis}
\affiliation{KBR, Inc, NASA Ames Research Center, Moffett Field, California 94035, United States}
\affiliation{Materials Sciences Division, Lawrence Berkeley National Laboratory, Berkeley, California 94720, United States}
\author{Jonah B. Haber}
\affiliation{Department of Materials Science and Engineering, Stanford University, Stanford, California 94305, United States}
\affiliation{Materials Sciences Division, Lawrence Berkeley National Laboratory, Berkeley, California 94720, United States}
\author{Zhenglu Li}
\affiliation{Mork Family Department of Chemical Engineering and Materials Science, University of Southern California, Los Angeles, California 90089, USA}
\affiliation{Materials Sciences Division, Lawrence Berkeley National Laboratory, Berkeley, California 94720, United States}
\affiliation{Department of Physics, University of California Berkeley, Berkeley, United States}
\author{Christopher J. N. Coveney}
\affiliation{Department of Physics, University of Oxford, Oxford OX1 3PJ, United Kingdom}
\author{Steven G. Louie}
\affiliation{Materials Sciences Division, Lawrence Berkeley National Laboratory, Berkeley, California 94720, United States}
\affiliation{Department of Physics, University of California Berkeley, Berkeley, United States}
\author{Marina R. Filip}
\affiliation{Department of Physics, University of Oxford, Oxford OX1 3PJ, United Kingdom}
\author{Jeffrey B. Neaton}
\email{jbneaton@lbl.gov}
\affiliation{Materials Sciences Division, Lawrence Berkeley National Laboratory, Berkeley, California 94720, United States}
\affiliation{Department of Physics, University of California Berkeley, Berkeley, United States}
\affiliation{Kavli Energy NanoScience Institute at Berkeley, Berkeley, United States}

\date{\today}
\begin{abstract}
The properties of excitons, or correlated electron-hole pairs,
are of paramount importance to optoelectronic
applications of materials. A central component of exciton physics is the electron-hole interaction, which is commonly treated as screened solely by electrons within a material. However, nuclear motion
can screen this Coulomb interaction as well, with several recent studies developing model approaches for approximating the phonon screening to the properties of excitons. While these model approaches tend to
improve agreement with experiment for exciton properties, they rely on several approximations that restrict their
applicability to a wide range of materials, and thus far they have neglected
the effect of finite temperatures. Here, we develop a fully first-principles, parameter-free approach to compute the temperature-dependent effects of phonon screening within the \emph{ab initio} $GW$-Bethe Salpeter equation framework.
We recover previously proposed models of phonon screening as well-defined limits of our general framework, and discuss their validity 
by comparing them against 
our first-principles results.
We develop an efficient computational workflow and apply it
to a diverse set of semiconductors, specifically AlN, CdS, GaN, MgO and $\text{SrTiO}_3$.
We demonstrate under different 
physical scenarios how excitons may be screened by multiple polar optical or acoustic phonons, how their binding energies can exhibit strong temperature dependence, and the ultrafast timescales on which they dissociate into free electron-hole pairs.  

\end{abstract}

\maketitle

\section{Introduction}
Excitons, correlated electron-hole pairs often generated upon photoexcitation, are critical to semiconductor optoelectronic applications. The dissociation of bound excitons into
free charge carriers is central to photovoltaics~\cite{Grancini2013,Savenije2014}, while their ability to recombine and emit
light underpins applications such as light-emitting diodes~\cite{Ai2018,Reineke2009}. 
Capturing the materials-specific many-body interactions of electrons and holes in solids requires careful theoretical descriptions of screening and scattering mechanisms, and constitutes a major challenge for first-principles approaches. The state-of-the-art \emph{ab initio} framework to describe excitons within many-body perturbation theory is based on the \emph{GW} approximation~\cite{Lars1965,Hybertsen1986} and Bethe-Salpeter
equation (BSE)~\cite{PhysRevLett.80.4510, Benedict1998,Rohlfing1998,Rohlfing2000} (\emph{GW}+BSE), where $G$ is the one-particle Green's function and $W$ is the screened Coulomb interaction. A key ingredient responsible for the formation of bound
excitons is the frequency-dependent ($\omega$) screened Coulomb
interaction~\cite{Strinati1988,Rohlfing1998,Rohlfing2000}
\begin{equation}
\label{eq:screening}
    W(\mathbf{r},\mathbf{r}',\omega)=\int d\mathbf{r}''\epsilon^{-1}(\mathbf{r},\mathbf{r}'',\omega)v(\mathbf{r}'',\mathbf{r}'),
\end{equation}
where $v$ is the bare Coulomb interaction and $\epsilon(\mathbf{r},\mathbf{r}'',\omega)$ is the frequency-dependent, non-local dielectric function. 
Most commonly, $\epsilon$ is obtained within the random-phase approximation (RPA)~\cite{Ceperley1980} and 
describes screening only originating from the perturbation of the electron density in the limit of clamped ions, the \emph{electronic} screening.

However, exciton binding energies computed within the standard \emph{ab initio} \emph{GW}+BSE framework can overestimate experiments, by up to a factor of three in certain heteropolar crystals~\cite{Bokdam2016}, a discrepancy attributed to the screening 
of Coulomb interactions due to polar ionic vibrations~\cite{Bechstedt2019,Umari2018,Bokdam2016,Schleife2018,Fuchs2008,Park2022}, normally neglected in standard approaches. 
The phonon contribution to the 
low-frequency (static) dielectric constant, $\epsilon_0$, in addition to the high-frequency (optical) counterpart, $\epsilon_{\infty}$, quantifies the screening originating from ionic vibrations. This raises the question of which dielectric constant more appropriately describes the screening of weakly interacting electrons and holes (\emph{i.e.} with a binding energy comparable to phonon energies),
with several studies proposing an intermediate \emph{ad hoc} ``effective" dielectric constant with a value between these two limits~\cite{Mahanti1970,Mahanti1972,Herz2018}, in an effort to account for screening from both electrons and phonons and to capture experimental trends.
An alternative approach is to modify the dielectric function of Eq.\,\ref{eq:screening} so that it includes the
contributions from polar phonons~\cite{Umari2018,Adamska2021}.
Several model approaches have been reported that include phonon contributions $W^{ph}$ to the screened Coulomb interaction to complement the
standard clamped-ion electronic screening $W=W^{el}$~\cite{Haken1956,Haken1958,Mahanti1970,Filip2021}. The results
from such approaches can vary significantly depending on approximations that are used; they are limited to the description of screening from polar phonons; and they do not yield detailed microscopic insights into these effects relative
to electronic screening.
A rigorous, fully first-principles, description of both electronic and phonon screening of excitons on
equal footing is therefore necessary to develop truly predictive accuracy for exciton properties in
diverse materials, and to reach a deeper understanding of factors affecting the relative importance and interplay of these two effects. 

Within many-body perturbation theory and to the lowest order in the electron-phonon interaction, the phonon-screened Coulomb interaction is given by~\cite{Baym1961,Hedin1965,Giustino2017}
\begin{equation}
\label{eq:phonon_screening}
W^{ph}(\mathbf{r},\mathbf{r}',\omega) = \sum_{\mathbf{q},\nu} D_{\mathbf{q},\nu}(\omega)g_{\mathbf{q},\nu}(\mathbf{r})g^*_{\mathbf{q},\nu}(\mathbf{r}'),
\end{equation}
where $D_{\mathbf{q},\nu}(\omega)$ is the propagator of a phonon with branch index $\nu$ at wavevector $\mathbf{q}$, and $g_{\mathbf{q},\nu}$ is the electron-phonon
vertex. Ref.~\cite{Filip2021} used this expression to derive a $0$\,K correction to the electron-hole interaction term in the \emph{ab initio} Bethe-Salpeter equation due to
phonon screening, subsequently approximating it
for ``hydrogenic" excitons and long range  polar electron-phonon interactions described by the Fr\"{o}hlich model. For a number of systems 
where these approximations are well-justified, this notably
improved agreement with experiment for exciton binding energies. 

Here, we generalize the framework in Ref.~\cite{Filip2021} to finite temperatures and implement it within a fully \emph{ab initio} BSE approach, accounting for the effects of phonon screening to lowest order in the electron-phonon interaction, at the level of  phonon exchange. Ref.~\cite{Filip2021} is one of several recent papers~\cite{Bechstedt2019, Adamska2021} that, building on pioneering work of Hedin and Lundqvist~\cite{Hedin1970} and Strinati~\cite{Strinati1988} decades earlier, include the effects of phonon screening on exciton properties, within \emph{ab initio} many-body perturbation theory. Closely related recent works treat exciton-phonon interactions in a modern first-principles context, and using cumulants~\cite{Cudazzo2020}, using two-particle Green’s functions~\cite{Antonius2022} and using a linear-response real-time framework~\cite{Paleari2022}, have derived phonon-renormalized exciton properties in terms of the exciton-phonon vertex, analogous to the one used in the seminal work of Toyozawa~\cite{Toyozawa1958}. Some recent \emph{ab initio} calculations of exciton-phonon interactions in solids capture phenomena such as phonon-assisted absorption and luminescence~\cite{Paleari2019,Cannuccia2019,Lechifflart2023}, temperature-dependent exciton localization~\cite{Alvertis2023_2} and shifts of the exciton energy~\cite{Alvertis2020}, and report exciton-exciton scattering rates~\cite{Chen2020,Chan2023,Cohen2023}. Here, starting from Eq.~\ref{eq:phonon_screening}, we develop and implement a temperature-dependent complex correction to the standard clamped-ion BSE kernel, referred to in what follows as the \emph{phonon kernel}, $K_{ph}$. Although it is possible to recover or go beyond the exciton-phonon self-energy of Ref.~\cite{Antonius2022} starting from $W_{ph}$ by including other higher-order diagrams in the BSE~\cite{Cudazzo2020}, here we retain only the phonon exchange diagram of Ref.~\cite{Antonius2022} (equivalently Eq.~\ref{eq:phonon_screening}), a judicious, computationally efficient, and physical low-order approximation for semiconductor systems (Ref.~\cite{Haken1958}) for which free electron- and hole-polaron radii (as described by Fan-Migdal diagrams) are on the same order as exciton radii~\cite{Mahanti1970,Pollmann1977}. As we show in what follows, within the limits of perturbation theory, the real part of $K_{ph}$ in the exciton basis provides a quantitative prediction of the temperature-dependent renormalization of exciton  (binding) energies via phonon screening, and the imaginary part can lead to a quantitative rate of dissociation of an exciton into free electron and holes through absorption of a phonon within the approximations made here.

Demonstrating this in a select set of semiconductors, we
predict that phonons are responsible for a $50\%$ reduction of the
exciton binding energy of CdS at room temperature, an effect that is
highly temperature-dependent, with acoustic phonons having a substantial
contribution. Moreover, we predict that phonon absorption
by excitons in GaN contribute to their ultrafast dissociation into free charge carriers with a timescale that is consistent
with experimental measurements for similar materials. In $\text{SrTiO}_3$, we find multiple polar phonons can contribute to the screening of excitons at once, leading to a significant overall reduction of the exciton binding energy. Finally, we show how approximations to our first-principles results lead to 
models that are commonly used in the literature, such as the Haken 
potential~\cite{Haken1956,Haken1958}, and we discuss the validity of these models for different 
systems. 

The structure of this paper is as follows. Section\,\ref{first_principles} presents the theoretical background of our work and summarizes the derivation of the first-principles phonon correction to the BSE kernel at finite temperatures, while Sections\,\ref{real_part} and\,\ref{imaginary_part} focus on the real and imaginary parts of this kernel respectively, and the observables that may be extracted from these quantities. Appendix\,\ref{radii} summarizes exciton and polaron radii of the systems studied as a part of this work, as estimated within the Wannier-Mott and weakly coupled Fr\"{o}hlich polaron regime respectively,
which justifies certain approximations appearing in Section\,\ref{real_part}. Appendix\,\ref{Imkph} provides a more in-depth look into the imaginary part of the phonon kernel and its physical interpretation. In Section\,\ref{models} we examine different approximations to the phonon kernel, and
connect our work to model results from the literature.
In Section\,\ref{results} we present the bulk of our computational results. Specifically, Section\,\ref{computational_details} provides relevant
computational details, while Section\,\ref{comparison} presents our first-principles results at $0$\,K
for a range of systems, and compares these to the predictions of various models of phonon screening, also discussing the origin of observed differences. 
Section\,\ref{finite_temperatures} presents
an in-depth application of our \emph{ab initio} workflow to selected
materials and their temperature-dependent phonon screening
and exciton dissociation properties. For these selected systems, the
convergence of their phonon screening properties is demonstrated in detail 
in Appendix\,\ref{appendix_convergence}. For $\text{SrTiO}_3$ we compare the \emph{ab initio} and model results in greater detail in Appendix\,\ref{STO_comparison}. Additionally, Section\,\ref{acoustic} quantifies the importance of acoustic phonons in screening excitons and the importance of piezoelectric interactions. 
Finally in 
Section\,\ref{discussion}, we provide a discussion and outlook for our 
work.

\section{First-principles phonon kernel}
\label{first_principles}

The Bethe-Salpeter equation (BSE) within the Tamm-Dancoff approximation for zero-momentum excitons in reciprocal space is written as~\cite{Rohlfing2000}
\begin{align}
    \label{eq:BSE}
    (E_{c\mathbf{k}}-E_{v\mathbf{k}})A^S_{cv\mathbf{k}} +\sum_{c'v'\mathbf{k}'}\bra{cv\mathbf{k}}K^{eh}\ket{c'v'\mathbf{k}'}A^S_{c'v'\mathbf{k}'}\\ \nonumber =\Omega^SA^S_{cv\mathbf{k}},
\end{align}
where $E_{c\mathbf{k}}$ and $E_{v\mathbf{k}}$ are the quasiparticle energies of conduction and valence bands, respectively.
The BSE of Eq.\,\ref{eq:BSE} describes excited states  accounting only
for electronic screening effects, and will hence be referred to as the ``bare'' BSE, to distinguish it from the case where phonon screening is included.
The coefficients  $A^S_{cv\mathbf{k}}$ describe the corresponding excited state $S$ with excitation energy $\Omega_S$ as a linear combination of free electron-hole pair wave functions ($\ket{cv\mathbf{k}}$, typically obtained from a density functional theory (DFT) calculation), namely
\begin{equation}
    \label{eq:exciton}
    \ket{S}=\sum_{cv\mathbf{k}}A^S_{cv\mathbf{k}}\ket{cv\mathbf{k}}.
\end{equation}
The kernel $K^{eh}$ describes
the interaction between electrons and holes and consists of 
direct $(d)$ and exchange $(x)$ contributions, $K^{eh}=K^d+K^x$ . The repulsive exchange term $K^x$ depends on the bare Coulomb interactions and is frequency-independent. In the absence of spin-orbit coupling, this term is only non-zero for excitons of zero spin,
and is responsible for the different properties of singlet and
triplet excitons~\cite{Rohlfing2000}. On the other hand, the direct term is attractive, frequency-dependent, and involves the screened Coulomb interaction $W$. This term can be written as
\begin{align}
    \label{eq:direct_kernel}
    K^d_{cv\mathbf{k},c'v'\mathbf{k}'}(\Omega)=-\bra{cv\mathbf{k}}\frac{i}{2\pi}\int d\omega e^{-i\omega \delta}W(\mathbf{r},\mathbf{r}',\omega) \times  \\ \nonumber
    [\frac{1}{\Omega-\omega-\Delta_{c'\mathbf{k}'v\mathbf{k}}+i\delta}+\frac{1}{\Omega+\omega-\Delta_{c\mathbf{k}v'\mathbf{k}'}+i\delta}]\ket{c'v'\mathbf{k}'},
\end{align}
where $\delta$ is a small real and positive number and we have introduced the notation $\Delta_{c\mathbf{k}v\mathbf{k}}=E_{c\mathbf{k}}-E_{v\mathbf{k}}$. This kernel term is usually
computed in the clamped-ion limit including only electronic screening, \emph{i.e.} $W=W^{el}$~\cite{Rohlfing1998,Rohlfing2000}. While $W^{el}$ is fully frequency-dependent in principle, since exciton binding energies are much smaller than the band gaps and the plasmon energies in many insulators, the dynamical properties of $W^{el}$ are often neglected for weakly-bound excitons~\cite{Rohlfing2000}, namely $K^{el}_{cv\mathbf{k},c'v'\mathbf{k}'}(\Omega_S)=\bra{cv\mathbf{k}}W^{el}(\mathbf{r},\mathbf{r}',\omega=0)\ket{c'v'\mathbf{k}'}$. Dynamical effects can be important in some cases, for example in order to
account for free carrier screening from acoustic plasmons~\cite{Spataru2010,Zhang2023,Champagne2023}.

We now include the contribution of phonon
screening to the kernel, taking $W=W^{el}+W^{ph}$, which will yield a correction $K^{ph}$ to the direct kernel of Eq.\,\ref{eq:direct_kernel}. 
To obtain this phonon kernel $K^{ph}$, we introduce the phonon screening of Eq.\,\ref{eq:phonon_screening} into 
Eq.\,\ref{eq:direct_kernel}, with the phonon propagator appearing in Eq.\,\ref{eq:phonon_screening} written as~\cite{MahanGeraldD2013MP}
\begin{equation}
    \label{eq:propagator}
        D_{\mathbf{q},\nu}(\omega)=\frac{1}{\omega-\omega_{\mathbf{q},\nu}+i\delta}-\frac{1}{\omega+\omega_{\mathbf{q},\nu}-i\delta}.
\end{equation}
Therefore, as in Ref.~\cite{Filip2021}, we arrive to the following expression for the phonon kernel
\begin{align}
    \label{eq:phonon_kernel_0K}\small
    K^{ph}_{cv\mathbf{k},c'v'\mathbf{k}'}(\Omega)=-\sum_{\mathbf{q},\nu}g_{cc'\nu}(\mathbf{k}',\mathbf{q})g^*_{vv'\nu}(\mathbf{k}',\mathbf{q}) \times \\ \nonumber 
    \Bigg[\frac{1}{\Omega-\Delta_{c\mathbf{k}v'\mathbf{k}'}-\omega_{\mathbf{q},\nu}+i\eta}+\frac{1}{\Omega-\Delta_{c'\mathbf{k}'v\mathbf{k}}-\omega_{\mathbf{q},\nu}+i\eta}\Bigg],
\end{align}
where $\eta$ a small real and positive number. Here $\mathbf{q}=\mathbf{k}-\mathbf{k}'$ and
$g_{nm\nu}(\mathbf{k}',\mathbf{q})=\bra{n\mathbf{k}'+\mathbf{q}}g_{\mathbf{q}\nu}\ket{m\mathbf{k}'}$, which can be computed, for example, via density functional perturbation theory (DFPT)~\cite{Baroni2001} or beyond, via $GW$ perturbation theory (GWPT)~\cite{Li2019}.

The result of Eq.\,\ref{eq:phonon_kernel_0K} is only valid at zero temperature. We extend the phonon kernel to finite temperatures via the Matsubara formalism. Here the integral of Eq.\,\ref{eq:direct_kernel} is analytically continued into the complex plane and evaluated at imaginary bosonic Matsubara frequencies. Following this
well-established procedure~\cite{MahanGeraldD2013MP},
we obtain the following expression for the temperature-dependent phonon contribution to
the kernel
\begin{align}
    \label{eq:phonon_kernel_eh}\small
    K^{ph}_{cv\mathbf{k},c'v'\mathbf{k}'}(\Omega,T)=-\sum_{\mathbf{q},\nu}g_{cc'\nu}(\mathbf{k}',\mathbf{q})g^*_{vv'\nu}(\mathbf{k}',\mathbf{q}) \times
    \\ \nonumber \Bigg[\frac{N_B(\omega_{\mathbf{q},\nu},T)+1+N_B(\Delta_{c\mathbf{k}v'\mathbf{k}'},T)}{\Omega-\Delta_{c\mathbf{k}v'\mathbf{k}'}-\omega_{\mathbf{q},\nu}+i\eta}+\\ \nonumber\frac{N_B(\omega_{\mathbf{q},\nu},T)+1+N_B(\Delta_{c'\mathbf{k}'v\mathbf{k}},T)}{\Omega-\Delta_{c'\mathbf{k}'v\mathbf{k}}-\omega_{\mathbf{q},\nu}+i\eta}+\\ \nonumber
    \frac{N_B(\omega_{\mathbf{q},\nu},T)-N_B(\Delta_{c\mathbf{k}v'\mathbf{k}'},T)}{\Omega-\Delta_{c\mathbf{k}v'\mathbf{k}'}+\omega_{\mathbf{q},\nu}+i\eta}+\\ \nonumber\frac{N_B(\omega_{\mathbf{q},\nu},T)-N_B(\Delta_{c'\mathbf{k}'v\mathbf{k}},T)}{\Omega-\Delta_{c'\mathbf{k}'v\mathbf{k}}+\omega_{\mathbf{q},\nu}+i\eta}\Bigg],
\end{align}
where $N_B$ is
the Bose-Einstein occupation factor at temperature $T$. As we are concerned with temperatures near room temperature and materials with band gaps that are large compared to phonon energies, we use the fact that $N_B(\Delta_{c\mathbf{k}v\mathbf{k}},T)<<N_B(\omega_{\mathbf{q},\nu},T)$ moving forward.

The first two terms within the bracket of Eq.\,\ref{eq:phonon_kernel_eh} describe the
contribution of phonon emission to the kernel, and these terms are
finite even at $0$\,K. The last two terms within the bracket of 
Eq.\,\ref{eq:phonon_kernel_eh} are due to the absorption of 
phonons, and are only non-zero at temperatures greater than zero.
In this work, we implement the {\it ab initio} phonon kernel (as a matrix) rewritten in the bare or unperturbed exciton basis as
\begin{align}
    \label{eq:phonon_kernel}
    K^{ph}_{S,S'}(\Omega,T)=-\smashoperator{\sum_{cv\mathbf{k}c'v'\mathbf{k}'\nu}}A^{S*}_{cv\mathbf{k}}g_{cc'\nu}(\mathbf{k}',\mathbf{q})g^{*}_{vv'\nu}(\mathbf{k}',\mathbf{q})A^{S'}_{c'v'\mathbf{k}'}\times  \\ \nonumber
    \Bigg[\frac{N_B(\omega_{\mathbf{q},\nu},T)+1}{\Omega-\Delta_{c\mathbf{k}v'\mathbf{k}'}-\omega_{\mathbf{q},\nu}+i\eta}+\frac{N_B(\omega_{\mathbf{q},\nu},T)+1}{\Omega-\Delta_{c'\mathbf{k}'v\mathbf{k}}-\omega_{\mathbf{q},\nu}+i\eta} \\ \nonumber +\frac{N_B(\omega_{\mathbf{q},\nu},T)}{\Omega-\Delta_{c\mathbf{k}v'\mathbf{k}'}+\omega_{\mathbf{q},\nu}+i\eta}+\frac{N_B(\omega_{\mathbf{q},\nu},T)}{\Omega-\Delta_{c'\mathbf{k}'v\mathbf{k}}+\omega_{\mathbf{q},\nu}+i\eta}\Bigg],
\end{align}
where the off-diagonal matrix elements of $K^{ph}_{S,S'}$ describe the extent to which the exciton-phonon scattering significantly changes the character of the excited state $S$. In the cases we study, off-diagonal contributions to the $K^{ph}$ matrix are negligible, and we therefore do not include them in the following discussions. It should also be noted that since here we work within many-body perturbation theory, this scheme might lead to a poorer descriptions of system with strong non-perturbative electron-phonon coupling. Moreover, multi-phonon processes, which are not captured here, can become significant near room temperature for certain systems~\cite{Lee2020}. 

As elaborated on in Section\,\ref{computational_details}, converging the phonon screening
properties
requires the sum of Eq.\,\ref{eq:phonon_kernel} to be computed
on a dense grid in reciprocal space. Wannier-Fourier interpolation can be utilized to greatly accelerate the calculation of electron-phonon matrix elements $g$ via DFPT on a dense grid~\cite{Giustino2007}. However, as Wannier interpolation introduces a gauge inconsistency between the values of $g$ and the exciton coefficients, it has not so far been possible to take advantage of Wannier-based techniques to compute exciton-phonon interactions from first-principles~\cite{Chen2020,Zhang2021,Zhang2022,Chan2023}.
To address this challenge~\cite{GAUGE}, we use the strategy of computing the exact same set of Wannier-interpolated wavefunctions from DFT on a fine k-grid, and use their plane-wave basis representation to perform BSE calculations with the BerkeleyGW code~\cite{Deslippe2012}, and their Wannier basis representation to obtain electron-phonon matrix elements on the same fine k-grid from DFPT using the EPW code~\cite{Ponce2016}. Consequently, the gauge consistency is naturally guaranteed. 
Our work combines for the first time Wannier interpolation methods 
for electron-phonon and electron-hole interactions from first principles. Not only does our approach greatly accelerate our calculations, but it gives us access 
to the study of systems such as $\text{SrTiO}_3$ and the nitrides GaN and AlN, which would otherwise be computationally prohibitive.

Overall, inclusion of phonon screening in Eq.\,\ref{eq:BSE} leads to the following generalized BSE
\begin{align}
    \label{eq:BSE_Kph}
    (E_{c\mathbf{k}}-E_{v\mathbf{k}})A^{S,ph}_{cv\mathbf{k}} \\ \nonumber+\sum_{c'v'\mathbf{k}'}\bra{cv\mathbf{k}}K^{eh}+K^{ph}(\omega,T)\ket{c'v'\mathbf{k}'}A^{S,ph}_{c'v'\mathbf{k}'}\\ \nonumber =\Omega^{S,ph}A^{S,ph}_{cv\mathbf{k}}.
\end{align}
The superscript in the eigenvalues and eigenvectors $\Omega^{S,ph},A^{S,ph}$ highlights that the excited states $S$
arising from the solution of Eq.\,\ref{eq:BSE_Kph} now include the effect
of phonon screening. While one can diagonalize the combined kernel $K=K^{eh}+K^{ph}$ to find the solutions of Eq.\,\ref{eq:BSE_Kph},
for the systems considered in this work it is an excellent approximation to consider the effect of phonon screening as a small perturbation. We will therefore obtain the effect of phonon screening
on excited states within first-order perturbation theory in this work, as discussed
in the following Section\,\ref{real_part}.

\subsection{Real part of the phonon kernel}
\label{real_part}

For non-degenerate excitons of energy $\Omega_S$, the correction to their energies from phonon screening within first-order perturbation theory is
\begin{equation}
    \label{eq:real_part_kph}
    \Delta \Omega_S=\text{Re}\Bigg[\bra{S}K^{ph}(\Omega_S)\ket{S}\Bigg].
\end{equation}
For a subspace of degenerate excitons with dimension $N_S$, the phonon screening correction to the exciton energies is taken to be equal to the normalized trace of the $N_S \times N_S$ $K^{ph}$ matrix, calculated as $\Delta \Omega_S = \frac{1}{N_S}\text{Tr}[K^{ph}(\Omega_S)]$, which is gauge-invariant. Higher-order
corrections to the exciton energy due to phonon screening at the level of Eq.\,\ref{eq:phonon_screening} are unimportant, since these are proportional to the off-diagonal matrix elements of $K^{ph}_{S,S'}$, which for all systems studied here are of the order of $10^{-3}-10^{-2}$\,meV and can be safely neglected. For the same reason, first-order corrections
to exciton wavefunctions are negligible, and we therefore focus
in what follows on computing the correction of Eq.\,\ref{eq:real_part_kph} to the exciton energies.

We note here that in principle the effect of phonon screening should also be included in the $GW$ self-energy, yielding a correction $iGW^{ph}$ to the
quasiparticle band structure, and $E_{c\mathbf{k}},E_{v\mathbf{k}}$ in Eq.\,\ref{eq:BSE_Kph}. This frequency- and temperature-dependent correction is equivalent to the one described by the so-called Fan-Migdal self-energy~\cite{Cudazzo2020}, $\Sigma^{FM}$, the real part of which captures low-order energy renormalization and mass enhancement due to
electron-phonon interactions~\cite{Giustino2007}. Therefore, if one
were to include the effect of phonon screening both in the
$GW$ self-energy and in the BSE kernel this would generalize the BSE of Eq.\,\ref{eq:BSE_Kph} so that the bare electron and hole energies are substituted by
the respective polaron energies $\Tilde{E}_{c\mathbf{k}}=E_{c\mathbf{k}}+\Sigma_{c}^{FM}(\omega,T)$ and $\Tilde{E}_{v\mathbf{k}}=E_{v\mathbf{k}}+\Sigma_{v}^{FM}(\omega,T)$. Ref.~\cite{Marini2008} included such a temperature dependence for the quasiparticle energies, however did not account for the contribution of phonon screening to excitons. 

The exciton-phonon interaction was described in Ref.~\cite{Antonius2022} as the sum of three distinct contributions to the self-energy: the dynamical Fan-Migdal term, the dynamical phonon exchange term, and the frequency-independent Debye-Waller term. We note that the phonon exchange term in Ref.~\cite{Antonius2022} is equivalent to Eq.\,\ref{eq:phonon_kernel_eh} and that the Debye-Waller term does not affect the exciton binding energy. Thus, inclusion of the phonon kernel and the Fan-Migdal term in the BSE would be consistent with Ref.~\cite{Antonius2022}. Our approach is distinct from that of Ref.~\cite{Paleari2022}, which presented an alternative derivation of the exciton-phonon self-energy, based on a BSE in which an optical response function is defined as the variation of the electronic density with respect to the total potential (rather than only the external potential, as in Ref.~\cite{Antonius2022}); the use of this alternative approach was reported to lead to small differences in computed exciton linewidths for monolayer $\text{MoS}_2$ and $\text{MoSe}_2$~\cite{Paleari2022}. While a deeper comparison of the distinct approaches of Ref.~\cite{Antonius2022} and Ref.~\cite{Paleari2022} is reserved for future work, we expect that, quantitatively, changes to the exciton-phonon interaction originating from different formulations of the BSE will be small compared with the effects of phonon screening, given our \emph{ab initio} results in Section~\ref{results} and the results in Ref.~\cite{Paleari2022}. Moreover, we note that the exciton-phonon self-energy of Ref.~\cite{Paleari2022} also contains a phonon exchange diagram that is equivalent to that of Ref.~\cite{Antonius2022} and central to our work here. 

Neglecting the Fan-Migdal term of the exciton-phonon self-energy (\cite{Antonius2022,Paleari2022}) in constructing $K_{ph}$ is acceptable here because, by itself, this term, which describes the induced lattice polarization around the exciton’s constituent electron and hole as if the two particles were independent, fails to capture any modifications of the lattice polarization when the electron and hole are bound together. As noted originally in Ref.~\cite{Mahanti1970,Pollmann1977} and reiterated recently in Ref.~\cite{Filip2021}, depending on the relative exciton and electron (hole) polaron radii, the lattice polarization associated with the polarons may interfere significantly, and even possibly cancel each other. For several semiconductors, including those studied in this work, exciton and electron (hole) polaron radii have similar estimated values (see Appendix\,\ref{radii}), strongly suggesting that naive inclusion of the Fan-Migdal term at lowest order while ignoring interference effects that enter at higher order will lead to significant errors. We therefore restrict our focus to phonon exchange, and reserve a more rigorous treatment of polaronic mass enhancement and interference effects (\cite{Mahanti1970,Pollmann1977}) for future work. Thus, in what follows, the phonon screening shift of the lowest bound exciton energy, $\Delta \Omega_S$, as obtained through Eq.\,\ref{eq:real_part_kph}, is given by the reduction of the exciton binding energy of the same magnitude, \emph{i.e.} $\Delta E_B=-\Delta \Omega_S$.

\subsection{Imaginary part of the phonon kernel}
\label{imaginary_part}

\begin{figure}[tb]
    \centering
    \includegraphics[width=0.9\linewidth]{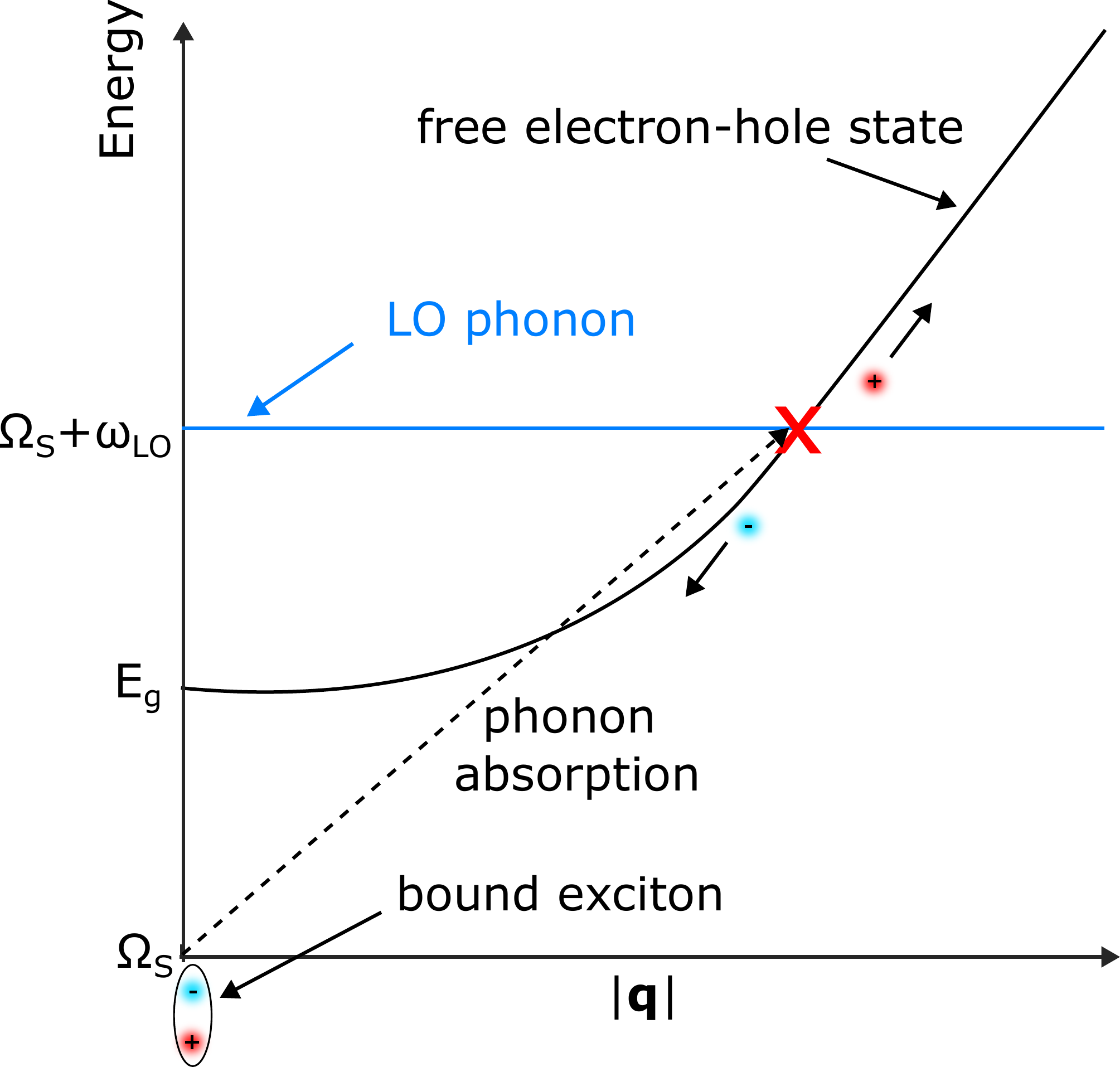}
    \caption{Schematic of the exciton dissociation process described by Eq.\,\ref{eq:lifetime} when a material with a band gap of $E_g$ and exciton energy $\Omega_S$ absorbs a longitudinal optical (LO) phonon of frequency $\omega_{LO}$, and the condition $\Omega_S+\omega_{LO}>E_g$ is satisfied.}
    \label{fig:dissociation_schematic}
\end{figure}

The phonon kernel of Eq.\,\ref{eq:phonon_kernel} is a complex quantity. 
For phonon emission and for $\omega=\Omega_S$, the imaginary part of the phonon kernel is proportional to 
\begin{equation}
\delta[\Omega_S-\omega_{\mathbf{q},\nu}-(E_{c\mathbf{k}'}-E_{v\mathbf{k}})]+\delta[\Omega_S-\omega_{\mathbf{q},\nu}-(E_{c\mathbf{k}}-E_{v\mathbf{k}'})],
\end{equation}
while for absorption it is proportional to
\begin{equation}
\delta[\Omega_S+\omega_{\mathbf{q},\nu}-(E_{c\mathbf{k}'}-E_{v\mathbf{k}})]+\delta[\Omega_S+\omega_{\mathbf{q},\nu}-(E_{c\mathbf{k}}-E_{v\mathbf{k}'})].
\end{equation}

For excitons of zero momentum, the conservation of energy condition, $\Omega_S-\omega_{\mathbf{q},\nu}=E_{c\mathbf{k}'}-E_{v\mathbf{k}}$,
which must be met for the emission channel to make non-zero 
contributions to the imaginary part of the phonon kernel, may only 
be satisfied for so-called ``resonant" excitons with energies greater than the quasiparticle band gap. This could occur for
indirect band gap materials with resonant excitons above the fundamental gap, \emph{e.g.} silver-pnictogen halide double perovskites~\cite{Biega2021}. Alternatively, this condition may be satisfied for higher-lying exciton states in the continuum, the analysis of which is beyond the scope of
the present study. On the other hand, phonon absorption, $\Omega_S+\omega_{\mathbf{q},\nu}=E_{c\mathbf{k}'}-E_{v\mathbf{k}}$, can occur if the energy of an exciton which has absorbed a phonon matches that of a free electron-hole pair, as schematically depicted in Fig.\,\ref{fig:dissociation_schematic} for the example of a longitudinal optical (LO) phonon with frequency $\omega_{LO}$. Specifically, this condition is only satisfied in a single-phonon process if the absorbed phonon has a greater or equal frequency than the exciton binding energy, \emph{i.e.} $\omega_{\mathbf{q},\nu}\geq E_B$.

The imaginary part of the exciton-phonon self-energy 
contains information about
the lifetime of excitons due to scattering from 
phonons into other excitons or free electron-hole pairs~\cite{Antonius2022}. As motivated in
Fig.\,\ref{fig:dissociation_schematic}, the phonon kernel
describes scattering from an initial exciton state to a
final free electron-hole pair, primarily due to phonon 
absorption. This process can be formally described within many-body perturbation theory and scattering theory by an $\mathcal{S}$-matrix of the following form~\cite{MahanGeraldD2013MP,Peskin1995}
\begin{align}
    \label{eq:s_matrix_born}
    \mathcal{S}=2\pi i\delta[\Omega_S+\omega_{\mathbf{q},\nu}- (E_{c\mathbf{k}}-E_{v\mathbf{k}'})]\times \\ \nonumber \bra{(c\mathbf{k},v\mathbf{k}'),N_B}K^{ph}\ket{S,N_B+1}.
\end{align}
The rate $\tau^{-1}_S$ of this scattering process described by the
$\mathcal{S}$-matrix of Eq.\,\ref{eq:s_matrix_born} 
is obtained by taking the time derivative of the square magnitude of this expression. Using the fact that $\delta^2[\Omega_S+\omega_{\mathbf{q},\nu}- (E_{c\mathbf{k}}-E_{v\mathbf{k}'})]=\lim_{t \rightarrow \infty}\frac{t}{2\pi}\delta[\Omega_S+\omega_{\mathbf{q},\nu}- (E_{c\mathbf{k}}-E_{v\mathbf{k}'})]$, we arrive at
\begin{align}
    \label{eq:dissociation_rate}
    \tau^{-1}_S\approx2\pi \sum_{c\mathbf{k}v\mathbf{k}'} \delta[\Omega_S+\omega_{\mathbf{q},\nu}- (E_{c\mathbf{k}}-E_{v\mathbf{k}'})]\times \\ \nonumber |\bra{(c\mathbf{k},v\mathbf{k}'),N_B}K^{ph}\ket{S,N_B+1}|^2,
\end{align}
in atomic units, where the approximately equal symbol is used because of making the
Born approximation, as detailed in Appendix\,\ref{Imkph}.
Eq.\,\ref{eq:dissociation_rate} is the same expression as that obtained through the
application of Fermi's golden rule for exciton dissociation, from an initial bound exciton
($\ket{S,N_B+1}$) to a final free electron-hole ($\ket{(c\mathbf{k},v\mathbf{k}'),N_B}$) state. This is analogous to the elementary treatment of the photoelectric effect in the hydrogen atom~\cite{HechtK.T2012QM}. 

Defining this scattering process through the $\mathcal{S}$-matrix of Eq.\,\ref{eq:s_matrix_born} generally requires the initial and final states
to be orthogonal to each other~\cite{Perea-Causin2021}. We can ensure this is the case within the Born approximation using the theory of rearrangement collisions~\cite{Lippmann1956,Sunakawa1960,Day1961}, as discussed in more detail in Appendix\,\ref{Imkph} and to be presented in Ref.~\cite{scattering}.
Employing the optical theorem for the  $\mathcal{S}$-matrix of Eq.\ref{eq:s_matrix_born}, the quantity $\text{Im}[K^{ph}_{SS}(\Omega_S,T)]$ can be shown to be equivalent to the rate of the exciton dissociation process depicted in Fig.\,\ref{fig:dissociation_schematic}, \emph{i.e.}
\begin{equation}
    \label{eq:lifetime}
    2|\text{Im}[K^{ph}_{SS}(\Omega_S,T)]|\approx\tau^{-1}_S(T).
\end{equation}
Thus, our framework enables the \emph{ab initio} calculation of exciton dissociation timescales for this particular channel for cases in which phonons with $\omega_{\mathbf{q},\nu}>E_B$ dominate. In this regime $\omega_{\mathbf{q},\nu}>E_B$ exciton dissociation competes with phonon-mediated exciton-exciton scattering, while exciton-exciton scattering dominates when $\omega_{\mathbf{q},\nu}<E_B$.
In what follows below, we will discuss GaN, a system
with ultra-fast exciton dissociation. 

\section{Approximations to the real part of the phonon kernel}
\label{models}
Having established our first-principles formalism to compute the phonon kernel, we now explore several common approximations to the real part of this quantity, \emph{i.e.} the perturbative correction $\Delta E_B$ to the exciton binding energy due to phonon screening, at $T=0$\,K.
The discussion of these approximations to $\Delta E_B$
reveals key physical
intuition and connects our work to previous studies. Since for this analysis we restrict ourselves to zero temperatures, we only
describe the effect of phonon emission on exciton binding energies.

\subsection{Fr\"{o}hlich electron-phonon coupling and hydrogenic excitons}
Several semiconductors of interest for optoelectronics, such as halide perovskites~\cite{Filip2021}, exhibit electron-phonon coupling dominated by long-range interactions with polar ionic vibrations, which can be described within the Fr\"{o}hlich model by the operator~\cite{doi:10.1080/00018735400101213}
\begin{equation}
    \label{eq:Frochlich}
    g^F_{\mathbf{q}}(\mathbf{r})=\frac{i}{|\mathbf{q}|}\Bigg[\frac{4\pi}{NV}\frac{\omega_{LO}}{2}\Bigg(\frac{1}{\epsilon_{\infty}}-\frac{1}{\epsilon_0}\Bigg)\Bigg]^{\frac{1}{2}}e^{i\mathbf{q}\cdot \mathbf{r}},
\end{equation}
where $N$ is the number of unit cells and $V$ the unit cell volume. 

Additionally, excitons of a wide class of materials behave in a hydrogenic manner
according to the Wannier-Mott limit~\cite{Wannier1937,TF9383400500}, with the reciprocal space wavefunction of their first excited state ($1s$) expressed as 
\begin{equation}
    \label{eq:hydrogenic}
    A_{\mathbf{k}}=\frac{(2a_o)^{3/2}}{\pi}\cdot \frac{1}{(1+a_o^2k^2)^2},
\end{equation}
where $a_o=1/(2E_B\mu)^{1/2}$ the exciton Bohr radius (in atomic units), $\mathbf{k}$ the
wavevector, and $\mu$ the exciton effective mass $1/\mu=1/m_e+1/m_h$,
with $m_e$ and $m_h$ the effective mass of the electron and hole, respectively.
Defining the exciton binding energy as $E_B=E_g-\Omega_S$, where $E_g$ is
the direct fundamental gap, assuming dispersive parabolic bands for the conduction and valence states, and ignoring the dispersion of the LO phonon, we arrive at the expression~\cite{Filip2021}:
\begin{align}
    \label{eq:kph_numerical}
    \Delta E_B = -\frac{8a_o^3}{\pi^2}\sum_{\mathbf{k}\mathbf{q}}\frac{|g_{\mathbf{q}\nu}|^2}{[1+a_o^2k^2]^2[1+a_o^2|\mathbf{k}+\mathbf{q}|^2]^2}\times \\ \nonumber [\frac{1}{E_B+\frac{k^2}{2m_e}+\frac{|\mathbf{k}+\mathbf{q}|^2}{2m_h}+\omega_{LO}}+\frac{1}{E_B+\frac{|\mathbf{k}+\mathbf{q}|^2}{2m_e}+\frac{k^2}{2m_h}+\omega_{LO}}].
\end{align}
In what follows, we compute this expression numerically for several systems on a grid of $\mathbf{k}-/\mathbf{q}-$points and we term the corresponding correction as $\Delta E^{\text{F-H}}_B$ (due to using the Fr\"{o}hlich and hydrogenic approximations), in order to differentiate it from the full \emph{ab initio} calculation. 

A way to further simplify Eq.\,\ref{eq:kph_numerical}, and derive analytic expressions for $\Delta E_B$, is to
take the limits $\mathbf{k}/\mathbf{q}\rightarrow \mathbf{0}$. In Sections\,\ref{Haken} and \ref{FHN}, we explore these limits.

\subsection{The $\mathbf{k}\rightarrow 0$ limit and the Haken potential}
\label{Haken}
Following Strinati~\cite{Strinati1988}, for excitons that are peaked around $\mathbf{k}=\mathbf{0}$, we can consider the $\mathbf{k}\rightarrow \mathbf{0}$ limit of Eq.\,\ref{eq:kph_numerical} and obtain
\begin{equation}
    \label{eq:Haken_exp}
    \Delta E_B = \bra{1s}V_{GH}(r)\ket{1s},
\end{equation}
where $\ket{1s}$ is the hydrogenic wavefunction and the potential $V_{GH}$ is given by
\begin{equation}
    \label{eq:VGH}
    V_{GH}(r)=v(r)\frac{\omega_{LO}}{2\epsilon_*(\omega_{LO}+E_B)}(e^{-r/\Tilde{r}_e}+e^{-r/\Tilde{r}_h}),
\end{equation}
with $v(r)$ the bare Coulomb potential and $\frac{1}{\epsilon_*}=(\frac{1}{\epsilon_{\infty}}-\frac{1}{\epsilon_0})$. 
The subscript of the potential term $V_{GH}$ indicates this is a generalized Haken potential, having the same form of the so-called Haken potential~\cite{Haken1956,Haken1958}, with the exception that $V_{GH}$ retains
the exciton binding energy, which is considered negligible in the derivation presented in  Ref.~\cite{Strinati1988}. 
We have defined the modified polaron radii for the electron and hole, compared to the usual definition~\cite{Mahanti1972}, as $\Tilde{r}_{e,h} = \frac{1}{\sqrt{2m_{e,h}(\omega_{LO}+E_B)}}$, retaining the exciton binding energy.
Therefore, the Haken potential
is trivially recovered as an approximation of our {\it ab initio} phonon screening expression, via a simplification of Eq.\,\ref{eq:kph_numerical}.

With the additional approximation $m_e=m_h$,
the expectation value for the shift of the exciton energy due to phonon
screening can be expressed in this limit as
\begin{equation}
    \label{eq:Haken_deb}
    \Delta E^{\mathbf{k}\rightarrow \mathbf{0}}_B=-\frac{2\omega_{LO}\Bigg(1-\frac{\epsilon_{\infty}}{\epsilon_0}\Bigg)}{\Bigg(1+\frac{\omega_{LO}}{E_B}\Bigg)\Bigg(1+\frac{1}{\sqrt{2}}\sqrt{1+\frac{\omega_{LO}}{E_B}}\Bigg)^2}.
\end{equation}

\begin{table*}[tb]
\centering
  \setlength{\tabcolsep}{6pt} 
\begin{tabular}{cccccc}
\hline
Material & Structure & $a$ ($\Ang$) & $c/a$ & Space Group & Identifier \\
\hline
AlN & Wurtzite & $3.128$ & $1.604$ &$\text{P6}_3\text{mc}$ & mp-661 \\
CdS &  Zincblende & $4.200$ & $1$ & $\text{F}\overline{4}3\text{m}$ & mp-2469 \\
GaN & Wurtzite & $3.215$ & $1.630$ & $\text{P6}_3\text{mc}$ & mp-804 \\
MgO & Halite, Rock Salt & $3.010$ & $1$ & $\text{Fm}\overline{3}\text{m}$ & mp-1265 \\
SrTi$\text{O}_3$ & Cubic Perovskite & $3.852$ & $1$ & $\text{Pm}\overline{3}\text{m}$ & mp-5229 \\
\hline
\end{tabular}
\caption{Studied materials, their structure, lattice parameters, space group, and identifier in the Materials Project database~\cite{Jain2013}. We performed geometry optimization for the atomic positions of these systems using DFT within the PBE exchange-correlation functional, keeping their lattice parameters fixed, with the exception of $\text{SrTiO}_3$, for which we used the local density approximation (LDA) and optimized both the atomic positions and its lattice parameter (the LDA has been discussed in the literature to yield more accurate results for structural properties of $\text{SrTiO}_3$ compared to PBE~\cite{Zhang2017}).}
\label{table:structures}
\end{table*}

\subsection{The $\mathbf{q}\rightarrow 0$ limit}
\label{FHN}
While the Haken potential is based on taking $\mathbf{k}\rightarrow 0$ 
with the justification of a highly localized exciton in reciprocal space, Ref.~\cite{Filip2021} instead retained the $\mathbf{k}$ dependence and considered the limit $\mathbf{q}\rightarrow 0$ for the potential term in the expectation value of Eq.\,\ref{eq:kph_numerical} for the shift of the
exciton binding energy. This is more justified in materials with highly
dispersive electronic bands, where setting the electronic momentum to zero
can constitute an oversimplification. For such systems, the momentum $\mathbf{q}$ of a phonon
may be considered negligible compared to that of an electron or a hole. Using this approximation we obtain
\begin{equation}
    \label{eq:FHN_exp}
    \Delta E_B = \bra{1s}V^{\mathbf{q}\rightarrow \mathbf{0}}(r)\ket{1s},
\end{equation}
where
\begin{align}
    \label{eq:FHN_potential}
    V^{\mathbf{q}\rightarrow \mathbf{0}}(r)=v(r)\frac{a_o^2\omega_{LO}}{\epsilon_*(\omega_{LO}+E_B)}\\ \nonumber \cdot [\frac{1}{a_o^2-b_o^2}-\frac{1}{r}\cdot \frac{2a_ob_o^2}{(a_o^2-b_o^2)^2}(1-e^{-(\frac{1}{b_o}-\frac{1}{a_o})r})],
\end{align}
with $a_o$ the exciton Bohr radius and $b_o=\sqrt{\frac{1}{2\mu(\omega_{LO}+E_B)}}$.
By setting $m_e=m_h$, the expectation value of Eq.\,\ref{eq:FHN_exp} is found to be
\begin{equation}
    \label{eq:FHN_deb}
    \Delta E^{\mathbf{q}\rightarrow \mathbf{0}}_B
    =-2\omega_{LO}\Bigg(1-\frac{\epsilon_{\infty}}{\epsilon_0}\Bigg)\frac{\sqrt{1+\frac{\omega_{LO}}{E_B}}+3}{\Bigg(1+\sqrt{1+\frac{\omega_{LO}}{E_B}}\Bigg)^3},
\end{equation}
the result derived in Ref.~\cite{Filip2021}.
We will see in Section\,\ref{results}
that this limit yields results which are generally significantly closer
to the full \emph{ab initio} value for the shift of the exciton energy
compared to the generalized Haken ($\mathbf{k}\rightarrow \mathbf{0}$) case.

\section{\emph{Ab initio} results for select systems}
\label{results}

\subsection{Computational Details}
\label{computational_details}

Table\,\ref{table:structures} summarizes the different systems studied in this work, the specific structure in which these are studied, their lattice parameter, as well as their space group and identifier in the Materials Project database~\cite{Jain2013}. For all studied materials with the exception of $\text{SrTiO}_3$, we start by performing a geometry optimization of their atomic positions, leaving the lattice parameters fixed. For this we employ DFT, as implemented within the Quantum Espresso software package~\cite{QE}, and we use the generalized gradient approximation (GGA) as formulated by Perdew, Burke and Ernzerhof (PBE)~\cite{pbe}. For $\text{SrTiO}_3$ we employ
the local density approximation (LDA)~\cite{Jones1989}
and optimize both the atomic positions and lattice parameters of this system; the LDA has been discussed in the literature to yield more accurate results for the phonon properties of $\text{SrTiO}_3$ compared to PBE~\cite{Zhang2017}. We then
employ DFPT using PBE, with the exception of $\text{SrTiO}_3$ for which we employ LDA to compute the
phonon dispersions of the studied materials on a $6\times 6\times 6$ grid of $\mathbf{q}$-points. 
Using the DFT-PBE Kohn-Sham wavefunctions (LDA for $\text{SrTiO}_3$) as a starting point, we perform $GW$ calculations as implemented within the BerkeleyGW code~\cite{Deslippe2012}, choosing the calculation parameters to converge the quasiparticle band gaps within $0.1$\,eV, following Refs.~\cite{Filip2021,Reyes-Lillo2016} and using a generalized plasmon pole model~\cite{Hybertsen1986} 
to compute the dielectric function at finite frequencies.
Specifically, we employ the following parameters for the $GW$ calculation: AlN ($400$ bands, $32$\,Ry polarizability cutoff, $6\times 6\times 6$ half-shifted k-grid), CdS ($500$ bands, $40$\,Ry polarizability cutoff, $6\times 6\times 6$ half-shifted k-grid), GaN ($400$ bands, $40$\,Ry polarizability cutoff, $4\times 4\times 4$ half-shifted k-grid), MgO ($600$ bands, $50$\,Ry polarizability cutoff, $6\times 6\times 6$ $\Gamma$-centered k-grid), $\text{SrTiO}_3$ ($1000$ bands, $14$\,Ry polarizability cutoff, $6\times 6\times 6$ half-shifted k-grid). 

The electronic BSE kernel is computed on the same $\mathbf{k}$-grid as the $GW$ eigenvalues, for three valence and
a single conduction band, with the exception of $\text{SrTiO}_3$, where nine valence bands and three conduction bands are employed instead.
For all cases, we use the patched sampling technique~\cite{Alvertis2023} to interpolate the kernel onto a patch drawn from a fine $100\times 100\times 100$ grid,  converging the size of the patch to ensure an accuracy of $1$\,meV or better for the exciton binding energy, as described in detail in Ref.~\cite{Alvertis2023}.
We find that a patch around $\Gamma$ with a crystal coordinate cutoff of $0.09$ is sufficient to converge the observables of interest in this study for all systems but MgO and $\text{SrTiO}_3$. For these two systems we extrapolate to the $N_k\rightarrow \infty$ limit by following the convergence rate of the value of $\Delta E^{\text{F-H}}_B$
obtained through the numerical integration of Eq.\,\ref{eq:kph_numerical} (see Appendix\,\ref{appendix_convergence}).
Moreover, we employ Wannier-Fourier interpolation~\cite{Giustino2007} in order to obtain the electron-phonon matrix elements $g_{mn\nu}(\mathbf{k},\mathbf{q})$
on the same fine patch for the $\mathbf{k}$- and $\mathbf{q}$-grid, using modified versions of the Wannier90~\cite{Pizzi2020} and EPW~\cite{Ponce2016} codes, in
order to ensure the gauge consistency of the electron-phonon matrix elements computed from EPW and the exciton coefficients computed using BerkeleyGW, using a workflow that will be described elsewhere~\cite{GAUGE}.

\subsection{Comparison between \emph{ab initio} and limiting cases at $0$\,K.}
\label{comparison}

\begin{table*}[tb]
\centering
  \setlength{\tabcolsep}{8pt} 
\begin{tabular}{ccccccccc}
\hline
system &  $E_B$ & $\omega_{LO}$ & $\epsilon_{\infty}$ & $\epsilon_0$ & $\Delta E_{B}^{ab\hspace{0.1cm} initio}$ & $\Delta E_{B}^{k\rightarrow 0}$ (Eq.\,\ref{eq:Haken_deb}) & $\Delta E_{B}^{q\rightarrow 0}$ (Eq.\,\ref{eq:FHN_deb}) & $\Delta E_{B}^{\text{F-H}}$ (Eq.\,\ref{eq:kph_numerical}) \\
\hline
GaN & $65$ & $87$ & $5.9$ & $10.8$ & $-15$ & $-6$ &$-22$ & $-15$\\
AlN & $143$ & $110$ & $4.5$& $8.7$ & $-29$ & $-16$ & $-36$ & $-29$ \\
MgO & $327$ & $84$ & $3.3$ & $11.3$ & $-46$  & $-26$ & $-52$ & $-48$\\
CdS & $39$ & $34$ & $6.2$ & $10.4$ & $-6$ & $-3$ & $-9$ & $-6$ \\
SrTi$\text{O}_3$ & $122$ & $98$ & $6.2$ &$409$ & $-44$ & $-25$ & $-65$ & $-51$ \\
\hline
\end{tabular}
\caption{Comparison of the shift of the exciton binding energy at $0$\,K due to phonon screening in the different studied systems. All energy values are given in meV. For reference, the computed value of the exciton binding energy $E_B$ as obtained from the solution of the bare BSE (without phonon effects) is given here, alongside the values computed within DFPT for the LO
phonon frequency and the high-/low-frequency dielectric constants $\epsilon_{\infty,o}$.}
\label{table:comparison}
\end{table*}

We start with the
\emph{ab initio} phonon screening correction to the exciton binding energy at $0$\,K, and
the values predicted through the various approximations outlined in Section\,\ref{models}. The results are summarized in Table~\ref{table:comparison}. 

Firstly, we note that for all studied systems, the \emph{ab initio} value
for the shift of the exciton binding energy due to phonon screening (as given by eq.\,\ref{eq:real_part_kph}) falls
between the values of the two limiting cases $\Delta E_{B}^{k\rightarrow 0}$ and $\Delta E_{B}^{q\rightarrow 0}$. The former of these limits, which corresponds to the well-known Haken potential, consistently underestimates
$\Delta E_B^{ab\hspace{0.1cm} initio}$, while the $q\rightarrow 0$ limit leads to a small overestimation. As introduced by Haken~\cite{Haken1956,Haken1958} and also elaborated by Strinati~\cite{Strinati1988}, exciton
coefficients in several semiconductors,
are highly localized around $\mathbf{k}=\mathbf{0}$ in reciprocal
space, motivating the $k\rightarrow 0$ approximation. However,
this approximation also suggests that one may neglect the dispersion of
electronic bands, only retaining the finite dispersion of the LO phonon. In materials such as the ones studied here, $A_{\mathbf{k}}$ assume appreciable non-zero values away from $\Gamma$, and this approximation is no longer valid, leading to the observed poor
agreement between $\Delta E_{B}^{k\rightarrow 0}$ and $\Delta E_{B}^{ab\hspace{0.1cm} initio}$. 
For the systems investigated in this work, considering the momentum of phonons to be negligible
compared to that of the electrons, \emph{i.e.} taking the $q\rightarrow 0$
in the energy denominators of Eq.\,\ref{eq:kph_numerical} and leading to
the expression of Eq.\,\ref{eq:FHN_deb}, is physically better justified, leading to better agreement with first-principles calculations. 

Moreover, the correction $\Delta E^{\text{F-H}}_B$, obtained through numerical integration of Eq.\,\ref{eq:kph_numerical}, which only
assumes hydrogenic excitons and an electron-phonon interaction governed by
the Fr\"{o}hlich vertex, is in excellent agreement with the first-principles results, for all systems but $\text{SrTiO}_3$, for reasons discussed in detail in Section\,\ref{finite_temperatures}. For the remaining systems studied here, the hydrogenic and Fr\"{o}hlich approximation are well-justified, as discussed
in Ref.~\cite{Filip2021}, leading to the excellent agreement between $\Delta E^{\text{F-H}}_B$ and $\Delta E_{B}^{ab\hspace{0.1cm} initio}$.

\subsection{First-principles calculations of finite temperature exciton binding energies and dissociation timescales}
\label{finite_temperatures}

In what follows, we will employ the fully first-principles phonon kernel and focus on temperature-dependent effects of phonon screening on excitons, in three of the materials selected from 
Table\,\ref{table:comparison}.
Specifically, we focus on CdS, the material with lowest LO phonon frequency, which indicates the potential
of this material for exhibiting substantial temperature-dependent phonon screening. Additionally, GaN is the only system studied here with $\omega_{LO}>E_B$, thus the absorption of an LO phonon from the exciton
may lead to dissociation of the electron-hole pair, according to Eq.\,\ref{eq:lifetime}. Finally, we discuss the effects of phonon screening on the cubic perovskite phase of $\text{SrTiO}_3$, which has a very large $\epsilon_0$ value, and is the only material showing significant deviations between the \emph{ab initio} and $\Delta E_B^{\text{F-H}}$ corrections to the exciton binding energy at $0$\,K. In Appendix\,\ref{thermal_expansion} we estimate the effects of thermal expansion on the phonon screening of excitons, which we generally find to be small. As also discussed in Appendix\,\ref{thermal_expansion}, thermal expansion can cause a modest reduction of the exciton binding energy, additional to that caused by phonon screening, leading to overall improved agreement with experiment. Nevertheless, phonon screening remains the dominant effect that determines the temperature dependence of exciton binding energies. 

\subsubsection{CdS}

\begin{figure}[tb]
    \centering
    \includegraphics[width=0.9\linewidth]{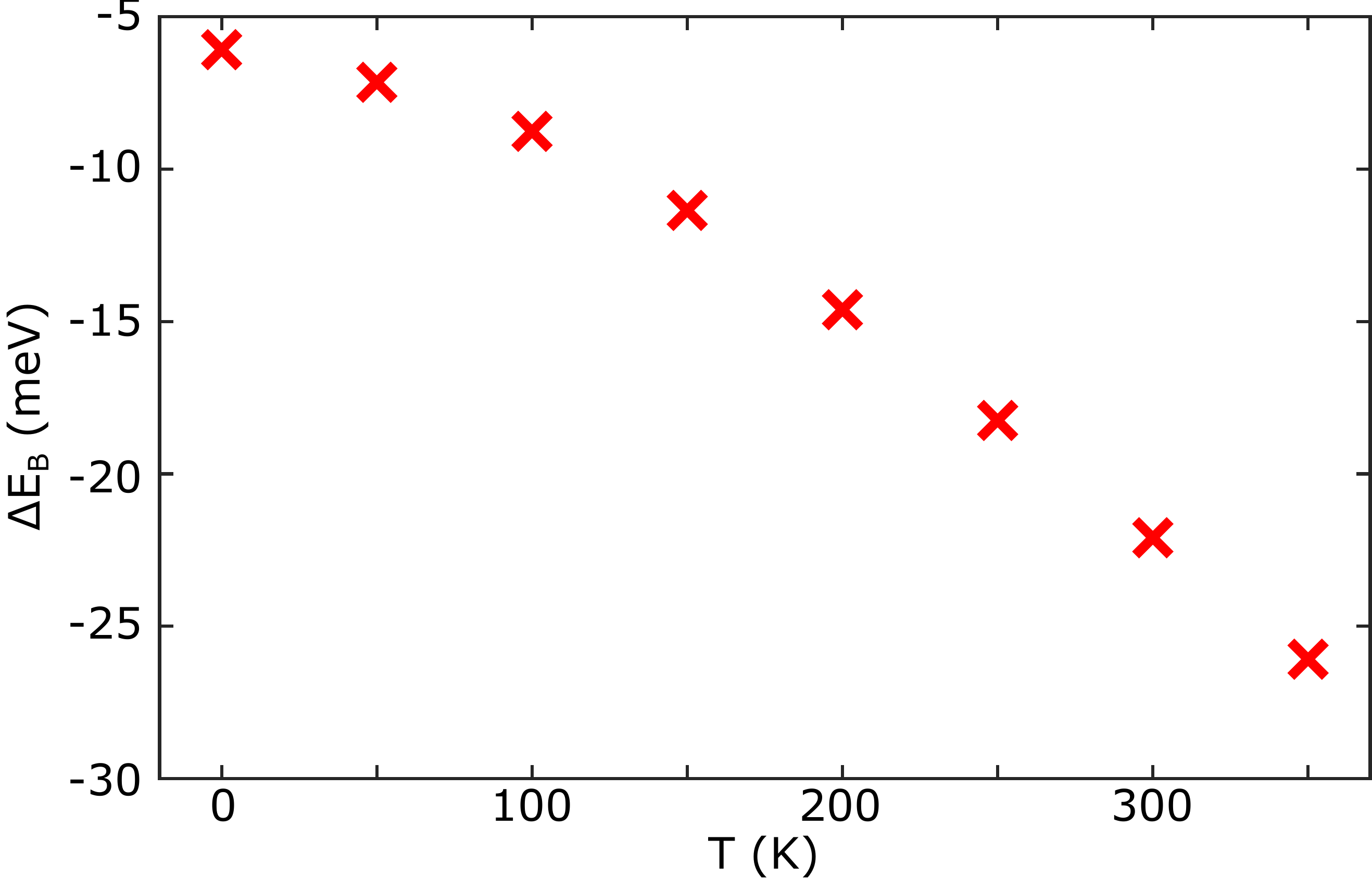}
    \caption{Calculated shift of the exciton binding energy of CdS due to phonon screening, as a function of temperature.}
    \label{fig:cds_deb_vs_T}
\end{figure}

\begin{figure}[tb]
    \centering
    \includegraphics[width=0.9\linewidth]{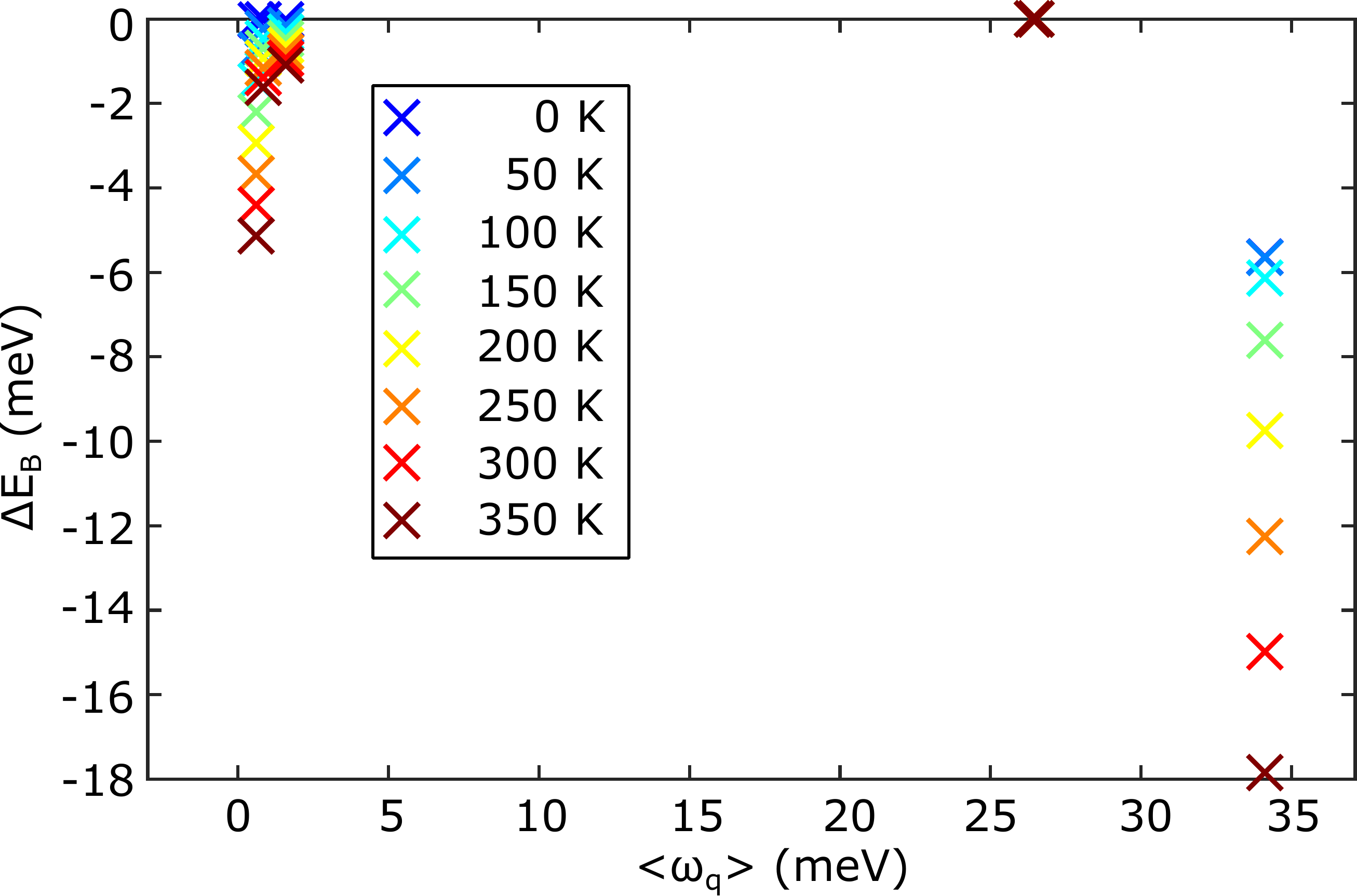}
    \caption{Contribution of different phonon modes to the exciton binding energy shift of CdS due to phonon screening, as a function of temperature. Here $\left\langle\omega_q\right\rangle$ denotes the average frequency of a particular phonon branch.}
    \label{fig:cds_deb_vs_omega}
\end{figure}

\begin{figure*}[tb]
    \centering
    \includegraphics[width=0.9\linewidth]{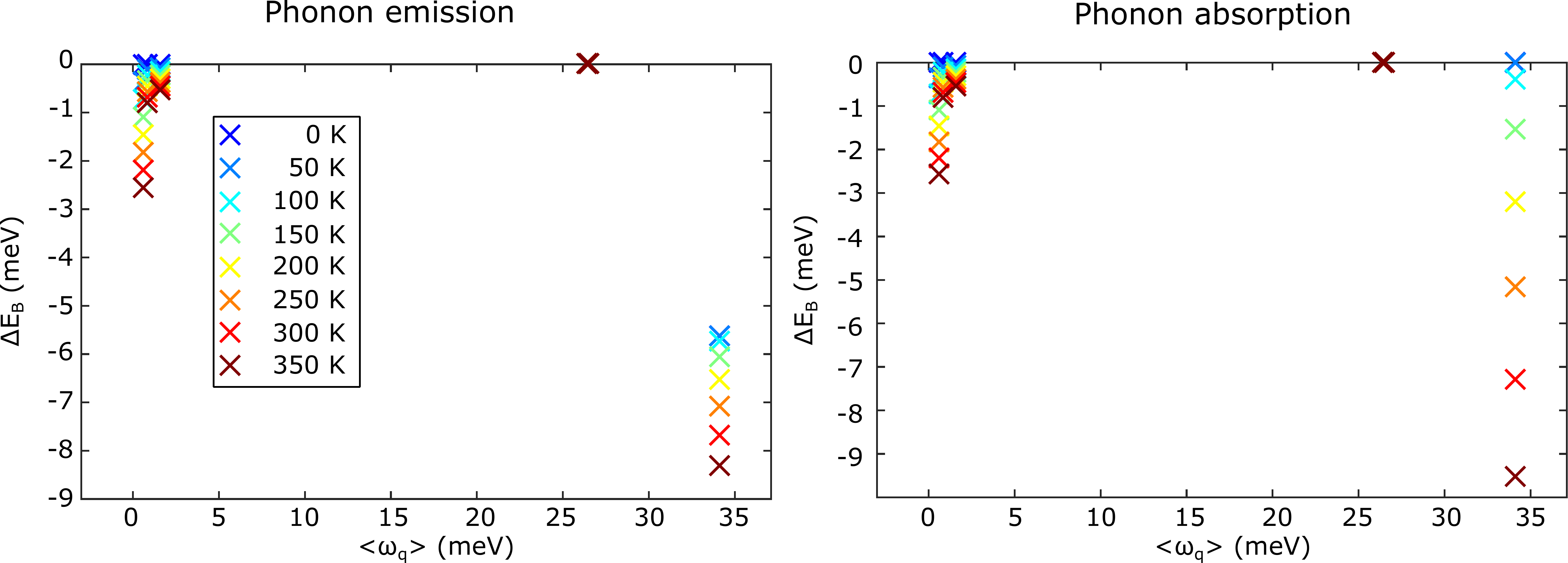}
    \caption{Contribution of emission and absorption of different phonon modes to the exciton binding energy shift of CdS due to phonon screening, as a function of temperature. Here $\left\langle\omega_q\right\rangle$ denotes the average frequency of a particular phonon branch.}
    \label{fig:cds_deb_vs_omega_em_abs}
\end{figure*}

\begin{figure}[tb]
    \centering
    \includegraphics[width=0.9\linewidth]{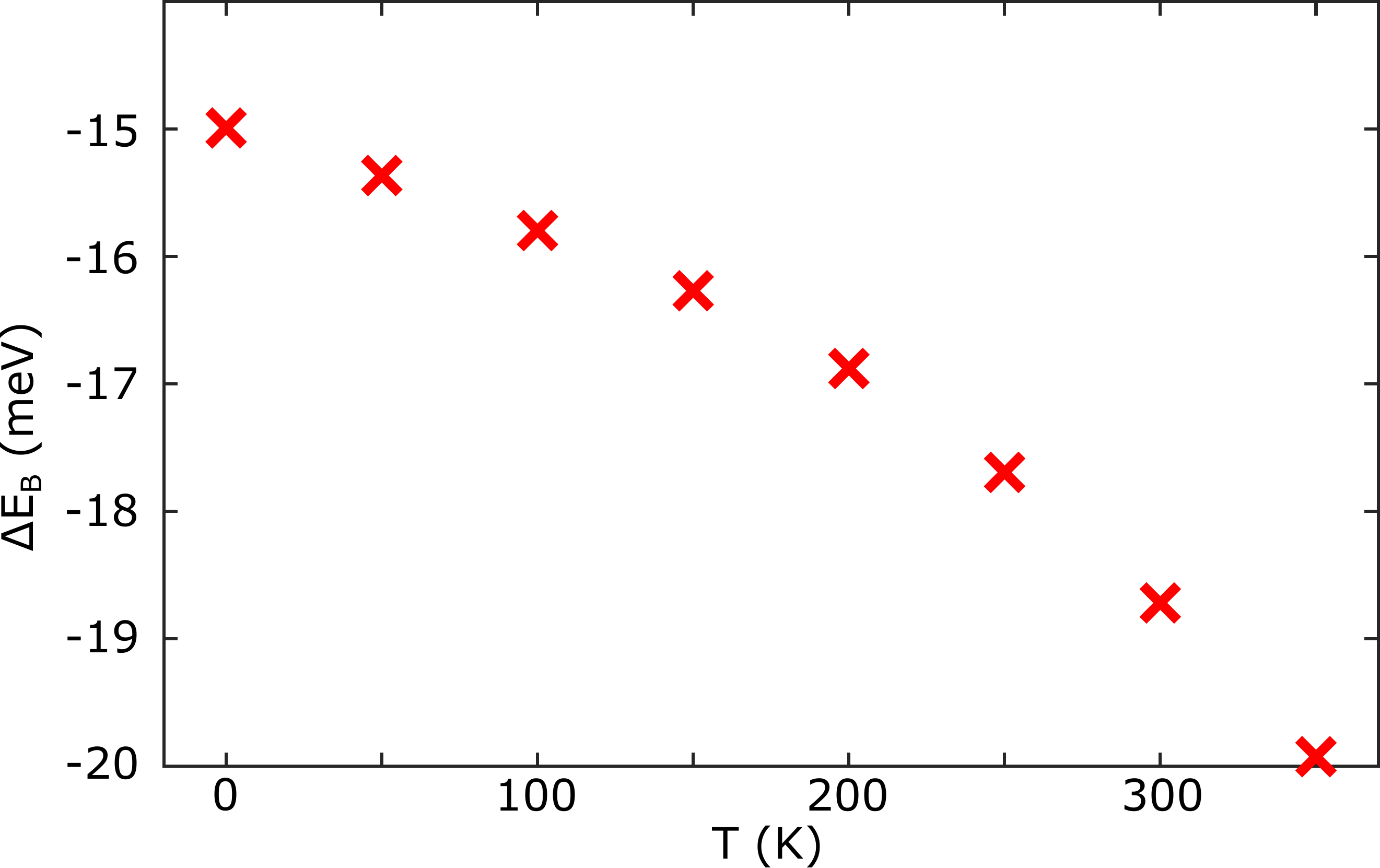}
    \caption{Calculated shift of the exciton binding energy of GaN due to phonon screening, as a function of temperature.}
    \label{fig:gan_deb_vs_T}
\end{figure}

\begin{figure}[tb]
    \centering
    \includegraphics[width=0.9\linewidth]{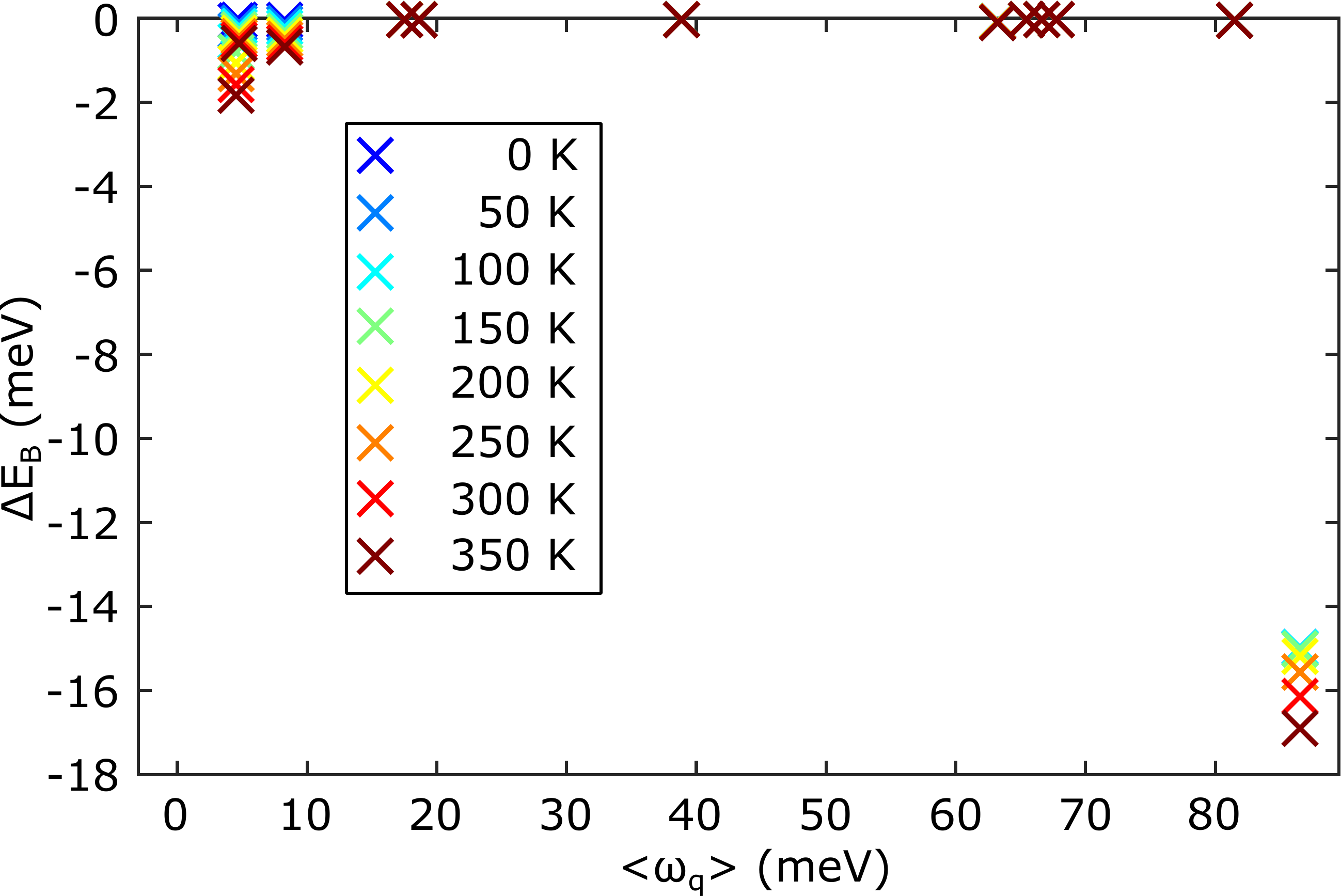}
    \caption{Contribution of different phonon modes to the exciton binding energy shift of GaN due to phonon screening, as a function of temperature. Here $\left\langle\omega_q\right\rangle$ denotes the average frequency of a particular phonon branch.}
    \label{fig:gan_deb_vs_omega}
\end{figure}

\begin{figure*}[tb]
    \centering
    \includegraphics[width=0.9\linewidth]{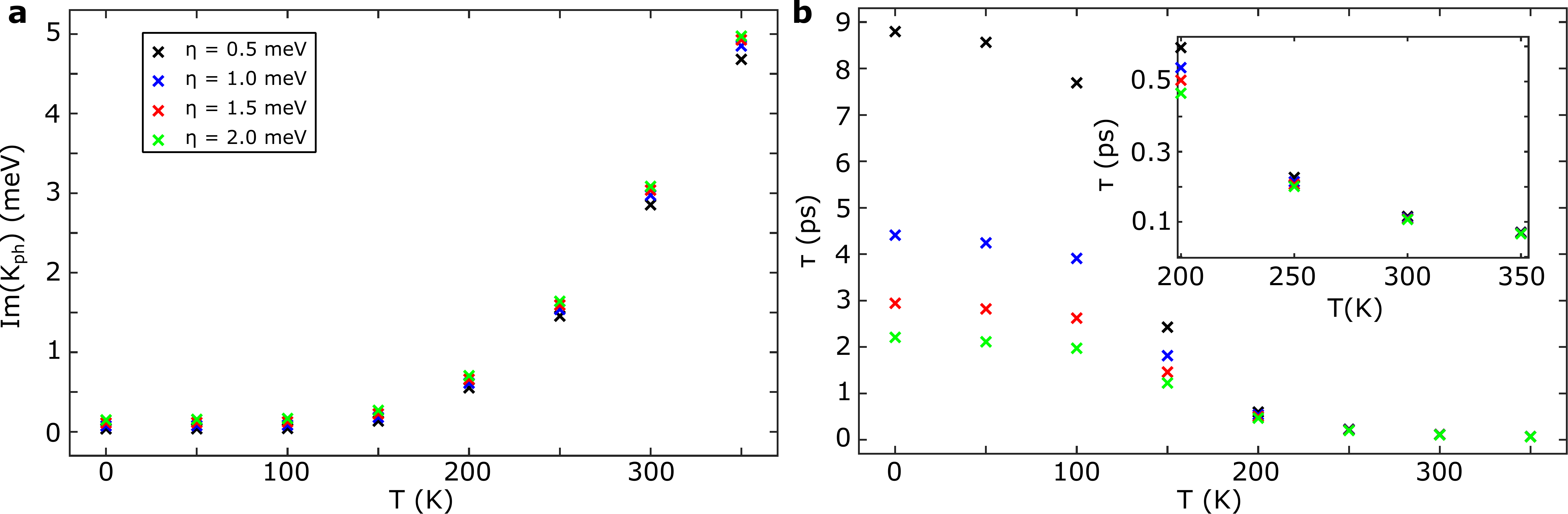}
    \caption{Imaginary part of the phonon kernel matrix elements between the $1s$ exciton basis states (panel \textbf{a}) and associated calculated timescale for exciton dissociation (panel \textbf{b}) as a function of temperature and the $\eta$ parameter for GaN. The inset of panel \textbf{b} provides a closer view of the range of temperatures around $300$\,K.}
    \label{fig:gan_Im_kph}
\end{figure*}

\begin{figure}[tb]
    \centering
    \includegraphics[width=0.9\linewidth]{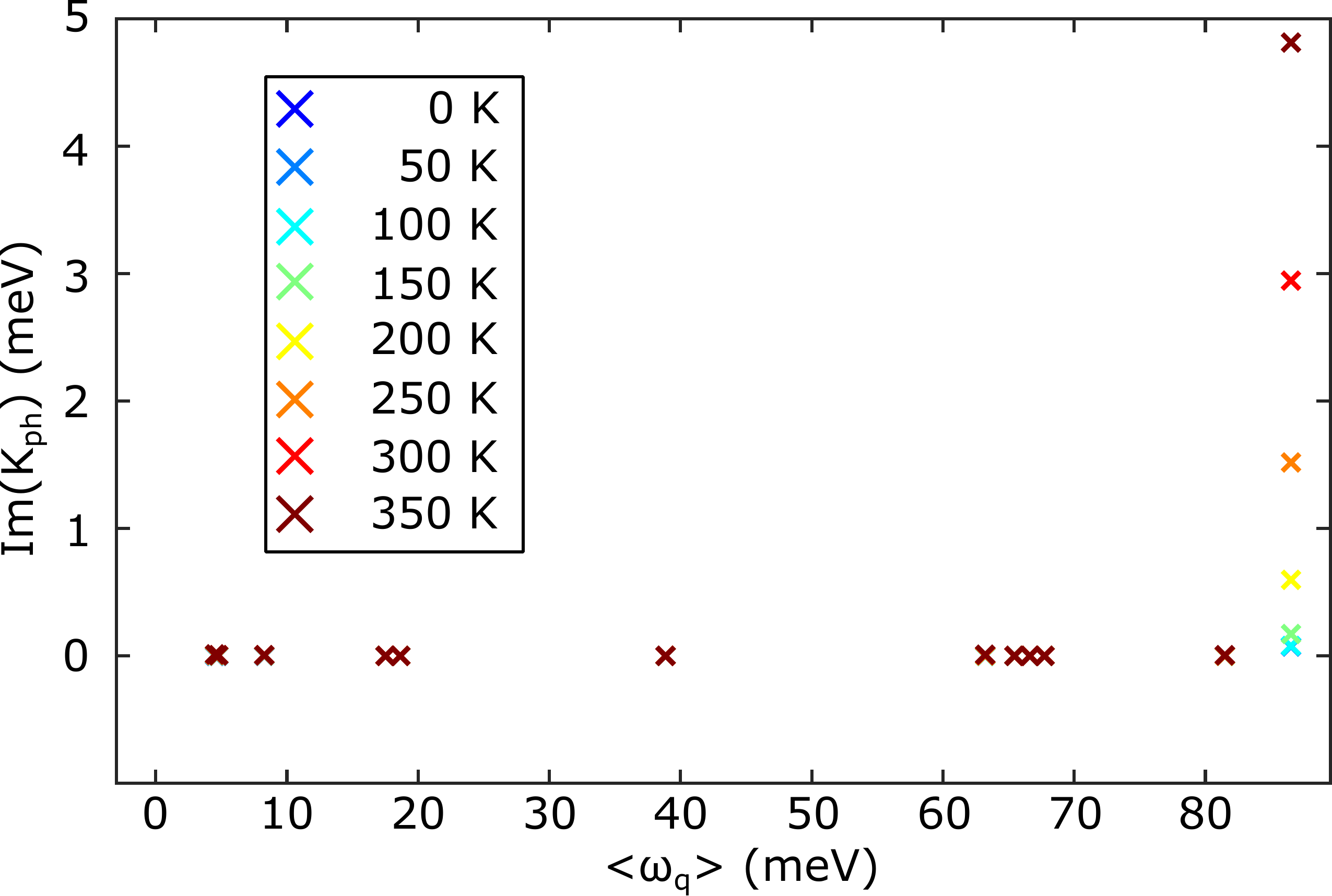}
    \caption{Contribution of different phonon modes to the imaginary part of the phonon kernel of GaN, as a function of temperature and for $\eta=1$\,meV. Here $\left\langle\omega_q\right\rangle$ denotes the average frequency of a particular phonon branch.}
    \label{fig:gan_imkph_vs_omega}
\end{figure}

In Fig.\,\ref{fig:cds_deb_vs_T}, we find that the correction of
the exciton binding energy of CdS due to phonon screening is strongly temperature-dependent. The low-temperature correction is equal to almost $-6$\,meV, and becomes more significant at room temperature, where it reaches a value of
$-22$\,meV. This suggests that the
bare, clamped-ion exciton binding energy of CdS of $39$\,meV as computed from BSE (see Table\,\ref{table:comparison}), will be renormalized by more than $50\%$, to
$17$\,meV at $300$\,K, due to phonon screening. The experimental value for the exciton binding energy has been reported at low
temperatures to be $28-30$\,meV~\cite{Voigt1979,Jakobson1994,Jeong2000}, in good agreement with our low-temperature result of a corrected exciton binding energy of $33$\,meV. Experimental temperature-dependent studies of CdS excitons assume the exciton binding energy to be temperature-independent~\cite{Jeong2000}, and extract it as the activation energy of a fit of photoluminescence data. Our results show that through the phonon modification of the BSE kernel, the excitonic interactions become themselves temperature-dependent, resulting in the strong temperature dependence of the exciton binding energy of CdS shown in Fig.\,\ref{fig:cds_deb_vs_T}.

It is instructive to decompose the computed phonon-induced screening correction to the exciton binding
energy into contributions from different phonon branches of CdS. As seen in Fig.\,\ref{fig:cds_deb_vs_omega}, unsurprisingly the vast majority of this effect is driven by the LO phonon with an average frequency of $34$\,meV across the Brillouin zone.
The contribution of this phonon to $\Delta E_B$ increases substantially with
temperature, due to its low frequency and high thermal activation. Interestingly, we also find that
at finite temperatures there is a non-negligible contribution of acoustic phonons to the screening, which is
strongly temperature-dependent due to the large thermal occupation factors of these modes. We return to the contribution of acoustic modes to phonon screening of excitons in Section\,\ref{discussion}.

Eq.\,\ref{eq:phonon_kernel} allows us to identify the
separate contributions of phonon absorption and emission to the phonon 
kernel. For CdS, we visualize in Fig.\,\ref{fig:cds_deb_vs_omega_em_abs}
the impact of these effects on the exciton binding energy. Notably, phonon emission is already active at
low temperatures, while phonon absorption only provides a minor contribution. This is intuitive, as the relevant phonons are unavailable to be absorbed from
the environment at low temperatures, according to Eq.\,\ref{eq:phonon_kernel}. The precise temperature where phonons become available to be absorbed may also depend on the level of theory used to obtain phonon frequencies (for example on the exchange-correlation functional~\cite{Hummer2009}  or the inclusion of anharmonic effects~\cite{Lazzeri2003}). Despite this expected sensitivity, we expect the contribution of the absorption term to $\Delta E_B$ at small temperatures will be small. On the other
hand, the phonon emission term survives even when $N_B(\omega_{\mathbf{q},\nu},T) \rightarrow 0$, and indeed the $0$\,K results of Table\,\ref{table:comparison} are entirely due to phonon emission. As the
temperature increases, the contribution of phonon absorption to the screening of the exciton becomes
more substantial, and eventually emission and absorption contribute equally.

\subsubsection{GaN}

The value of $\Delta E_B$ for GaN only shows weak temperature
dependence as seen in Fig.\,\ref{fig:gan_deb_vs_T}, since the frequency of its LO phonon has a value of $84$\,meV, significantly above room temperature. As for CdS, $\Delta E_B$ is dominated by the LO phonon (see Fig.\,\ref{fig:gan_deb_vs_omega}) while here too there is a small but non-negligible contribution from acoustic phonons.

Solution of the bare BSE gives an exciton binding energy of $E_B=65$\,meV for GaN, and by including the correction due to phonon
screening we predict it will decrease to $46$\,meV at $300$\,K. The
experimental values for the exciton binding energy of GaN are within
the range of $20-28$\,meV~\cite{Reimann1998,Muth1997}, with some studies measuring this
quantity at room temperature~\cite{Muth1997} and others at cryogenic temperatures~\cite{Reimann1998}. We were not able to find a systematic experimental study on the
temperature dependence of the exciton binding energy; however, it is clear
that while 
inclusion of phonon screening effects leads to better agreement with experiment, we still 
overestimate the exciton binding energy, similar to the case of the halide perovskites~\cite{Filip2021}. This result could be attributed to the
fact that the magnitude of electron-phonon interaction within DFPT might be underestimated compared to using
higher level theories~\cite{Yin2013,Monserrat2016,Li2019}, as well as
to the fact that we do not account for polaronic interference effects as
discussed in Section\,\ref{real_part}.

Among the systems studied in this work, GaN is the only one for which $\omega_{LO}>E_B$, making it a case where the absorption of a single LO
phonon by the exciton might lead to its dissociation into a free electron-hole pair. The fact that $\omega_{LO}>E_B$ manifests as a finite value of $\text{Im}[K^{ph}_{SS}(\Omega_S,T)]$, as shown in Fig.\,\ref{fig:gan_Im_kph}a. The value of
$\eta$ in Eq.\,\ref{eq:phonon_kernel} represents a small arbitrary broadening we
introduce to the energy levels appearing in the denominator, in order
to resolve the crossing between an exciton that has absorbed a phonon,
and the free electron-hole pair, as schematically shown in Fig.\,\ref{fig:dissociation_schematic}. Given the fine grid we are
employing here ($100\times 100 \times 100$) we find that values of $\eta$ within
the range of $0.5-2$\,meV are sufficient to satisfy this energy conservation condition. In Fig.\,\ref{fig:gan_Im_kph}a, we plot $\text{Im}[K^{ph}_{SS}(\Omega_S,T)]$ for a range of $\eta$ values within that window
and find that the change in the result is minor. Our values for $\text{Im}[K^{ph}_{SS}(\Omega_S,T)]$ are similar to the imaginary part of the self-energy of a model system~\cite{Antonius2022}, representing phonon-mediated exciton-exciton scattering, indicating that these effects are directly competing.

In Fig.\,\ref{fig:gan_Im_kph}b we plot the exciton dissociation
timescale for the lowest singlet exciton of GaN using Eq.\,\ref{eq:lifetime}, as a function of temperature and for different values of $\eta$. This exciton dissociation
process is entirely due to the absorption of LO phonons by the exciton,
as no other phonons contribute to the imaginary part of the phonon
kernel (see Fig.\,\ref{fig:gan_imkph_vs_omega}). We see from Fig.\,\ref{fig:gan_Im_kph}b that at low temperatures, the exciton
dissociation timescale varies significantly with changes in the value of
$\eta$. This is due to the fact that the imaginary part of the phonon
kernel for temperatures up to approximately $150$\,K assumes very small values of less than $0.5$\,meV, making even small changes in $\eta$ significant for its inverse in Eq.\,\ref{eq:lifetime}. Nevertheless
the exciton dissociation timescale $\tau$ becomes more stable with respect to changes in the value of $\eta$ at higher temperatures, as also highlighted in the inset of Fig.\,\ref{fig:gan_Im_kph}b. At $300$\,K we
find $\tau=111$\,fs. While we have not found experimental studies on time-resolved exciton dissociation in GaN to compare against,
it is encouraging that recent experiments employing ultrafast 2D electronic spectroscopy report exciton dissociation timescales that are similar to what we compute here, for semiconductors with
comparable exciton binding energies. Specifically, for GaSe an exciton dissociation timescale of $112$\,fs at room temperature has been reported~\cite{Allerbeck2021}, while for $\text{CH}_3\text{NH}_3\text{PbI}_3$, an exciton dissociation timescale of approximately $50$\,fs~\cite{Jha2018} was found.

It is also worth pointing out that the finite exciton lifetime described by the imaginary part of the phonon kernel, will manifest as a finite linewidth in absorption and emission spectra. However, exciton dissociation will only be one of several scattering processes contributing to the overall linewidth observed in experiment, with phonon-mediated exciton-exciton scattering~\cite{Chan2023,scattering},
Auger recombination~\cite{Bogardus1968}, and more, all contributing to the total linewidth. It is therefore no surprise that our value of approximately $3$\,meV for the imaginary part of the phonon kernel of GaN at $300$\,K is substantially smaller than the experimental linewidth of approximately $20$\,meV at the same temperature~\cite{Viswanath1998}.

\subsubsection{SrTi$\text{O}_3$}

\begin{figure}[tb]
    \centering
    \includegraphics[width=0.9\linewidth]{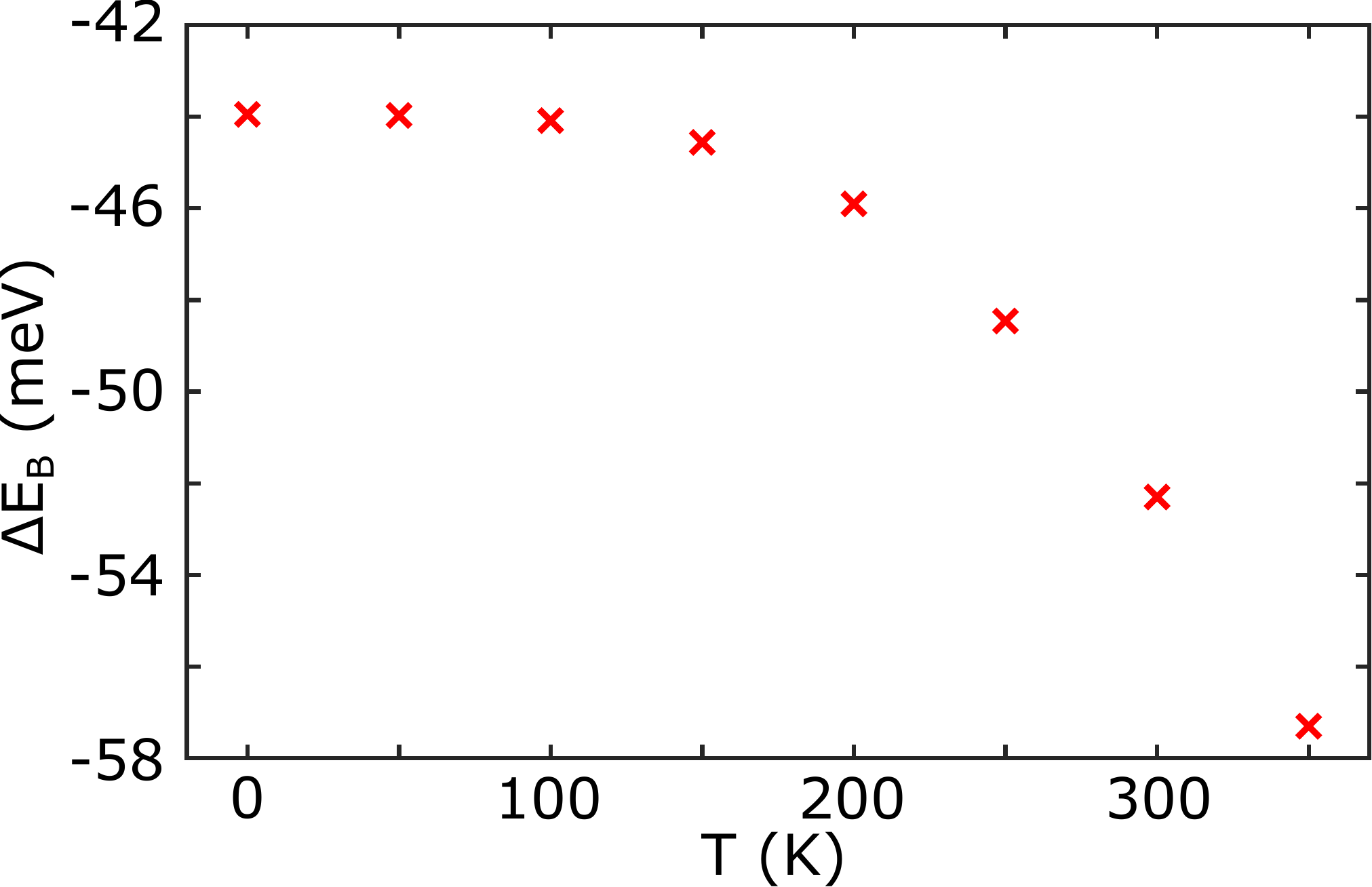}
    \caption{Calculated shift of the exciton binding energy of SrTi$\text{O}_3$ due to phonon screening, as a function of temperature.}
   \label{fig:STO_deb_vs_T}
\end{figure}

\begin{figure}[tb]
    \centering
    \includegraphics[width=0.9\linewidth]{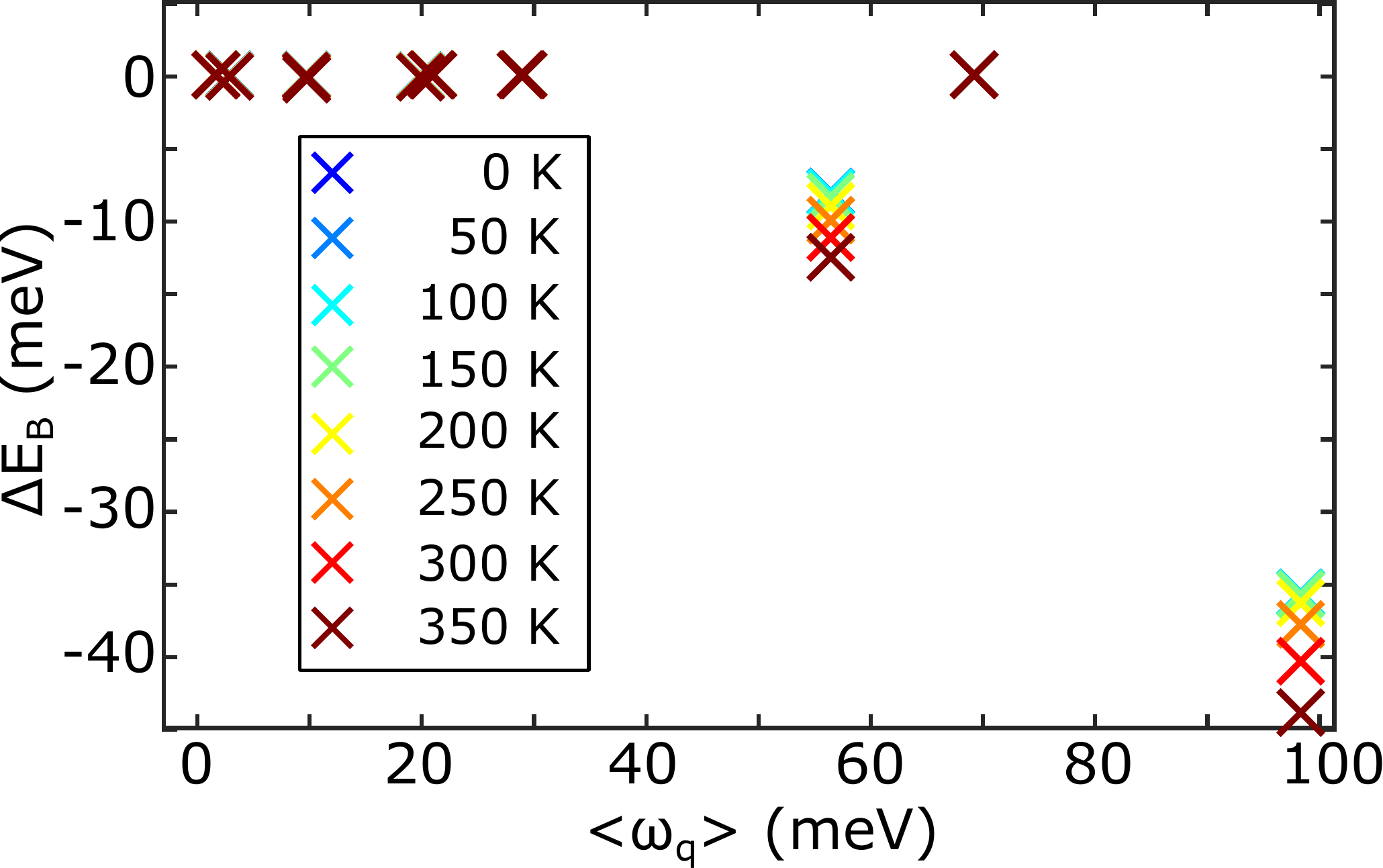}
    \caption{Contribution of different phonon modes to the exciton binding energy shift of SrTi$\text{O}_3$ due to phonon screening, as a function of temperature. Here $\left\langle\omega_q\right\rangle$ denotes the average frequency of a particular phonon branch.}
    \label{fig:STO_deb_vs_omega}
\end{figure}

Among the systems of Table\,\ref{table:comparison}, $\text{SrTiO}_3$
is the only one for which we find a substantial difference in the value
of $\Delta E_B^{\text{F-H}}$ at $0$\,K as predicted from the numerical integration of Eq.\,\ref{eq:kph_numerical}, and the full \emph{ab initio} correction $\Delta 
E_{B}^{ab\hspace{0.1cm} initio}$ to the exciton binding energy, following Eq.\,\ref{eq:phonon_kernel}. Moreover,
this is a system with a very large value for the low-frequency dielectric constant $\epsilon_0$, indicating a potentially very large contribution of phonons to the screened Coulomb interaction, making it particularly interesting for further study.

In Fig.\,\ref{fig:STO_deb_vs_T} we visualize the temperature-dependent
correction $\Delta E_B$ to the exciton binding energy for this system,
which we find to be equal to $-52$\,meV at $300$\,K, renormalizing the
bare clamped-ion exciton binding energy as computed within BSE by $43\%$, from $122$\,meV to $70$\,meV. By decomposing this correction
to the effect of individual phonons in Fig.\,\ref{fig:STO_deb_vs_omega},
we find that while phonon screening in this system is dominated by
the LO mode with a frequency of $98$\,meV (henceforth referred to as LO-1), there is a substantial contribution from an LO phonon with a lower
frequency of $57$\,meV (henceforth referred to as LO-2), which has been discussed previously to also exhibit
Fr\"{o}hlich-like coupling~\cite{Zhou2018}.

The reason behind the disagreement of the \emph{ab initio} and $\Delta E_B^{\text{F-H}}$ results
for the correction to the exciton binding energy is
that $\text{SrTiO}_3$ has multiple phonon modes contributing to the phonon 
screening and the static dielectric constant $\epsilon_0$. In these cases, the
Fr\"{o}hlich model, which is used in the numerical integration of Eq.\,\ref{eq:kph_numerical}, breaks down, and one needs to instead use the generalized Fr\"{o}hlich vertex of Ref.~\cite{Verdi2015}.
For a phonon $\nu$, this is written as
\begin{equation}
    \label{eq:long_range_g}
    g_{\mathbf{q},\nu}=i\frac{4\pi}{V}\sum_j \left(\frac{1}{2NM_j\omega_{\mathbf{q},\nu}}\right)^{1/2}\cdot \frac{\mathbf{q}\cdot \mathbf{Z}_j\cdot e_{\mathbf{j},\nu}(\mathbf{q})}{\mathbf{q}\cdot \epsilon_{\infty}\cdot\mathbf{q}},
\end{equation}
in atomic units. For atom $j$, $\mathbf{Z}_j$ is its Born effective charge tensor and $M_j$ its mass, while $V$ the unit cell volume, $N$ the number of unit cells, and $e_{\mathbf{j},\nu}(\mathbf{q})$ the phonon
eigenvectors. In 
Table\,\ref{table:STO_updated_model} we show that employing this generalized
Fr\"{o}hlich vertex in the numerical integration of Eq.\,\ref{eq:kph_numerical}
gives excellent agreement with the \emph{ab initio} result for the correction to the exciton binding energy of $\text{SrTiO}_3$ due to its LO-1 and LO-2 phonons. On the other hand, numerically integrating Eq.\,\ref{eq:kph_numerical} for the LO-1 and LO-2 phonons using the standard Fr\"{o}hlich vertex of Eq.\,\ref{eq:Frochlich} leads to significant discrepancies
with the \emph{ab initio} results.
The comparison between the case where we use the Fr\"{o}hlich and hydrogenic approximations, and the \emph{ab initio} result, is
discussed in more detail in Appendix\,\ref{STO_comparison}. 

\begin{table}[tb]
\centering
  \setlength{\tabcolsep}{6pt} 
\begin{tabular}{cccc}
\hline
phonon mode & $\Delta E_B^{\text{F-H}}$ & $\Delta E_{B,\text{gen.}}^{\text{F-H}}$ & $\Delta 
E_{B}^{ab\hspace{0.1cm} initio}$ \\
\hline
LO-1 & $-51$ & $-38$ & $-36$ \\
LO-2 & $-34$ & $-8$  & $-8$ \\
\hline
\end{tabular}
\caption{Phonon-resolved screening of the exciton binding energy of $\text{SrTiO}_3$ by the two LO modes of this material at $0$\,K (in meV). We compare full \emph{ab initio} level theory ($\Delta 
E_{B}^{ab\hspace{0.1cm} initio}$), the numerical integration of Eq.\,\ref{eq:kph_numerical} using the standard Fr\"{o}hlich vertex for each phonon ($\Delta E_B^{\text{F-H}}$), as well as the numerical integration of Eq.\,\ref{eq:kph_numerical} employing the generalized Fr\"{o}hlich vertex ($\Delta E_{B,\text{gen.}}^{\text{F-H}}$).}
\label{table:STO_updated_model}
\end{table}

\subsection{Phonon screening from acoustic modes}
\label{acoustic}

\begin{figure}[tb]
    \centering
    \includegraphics[width=0.9\linewidth]{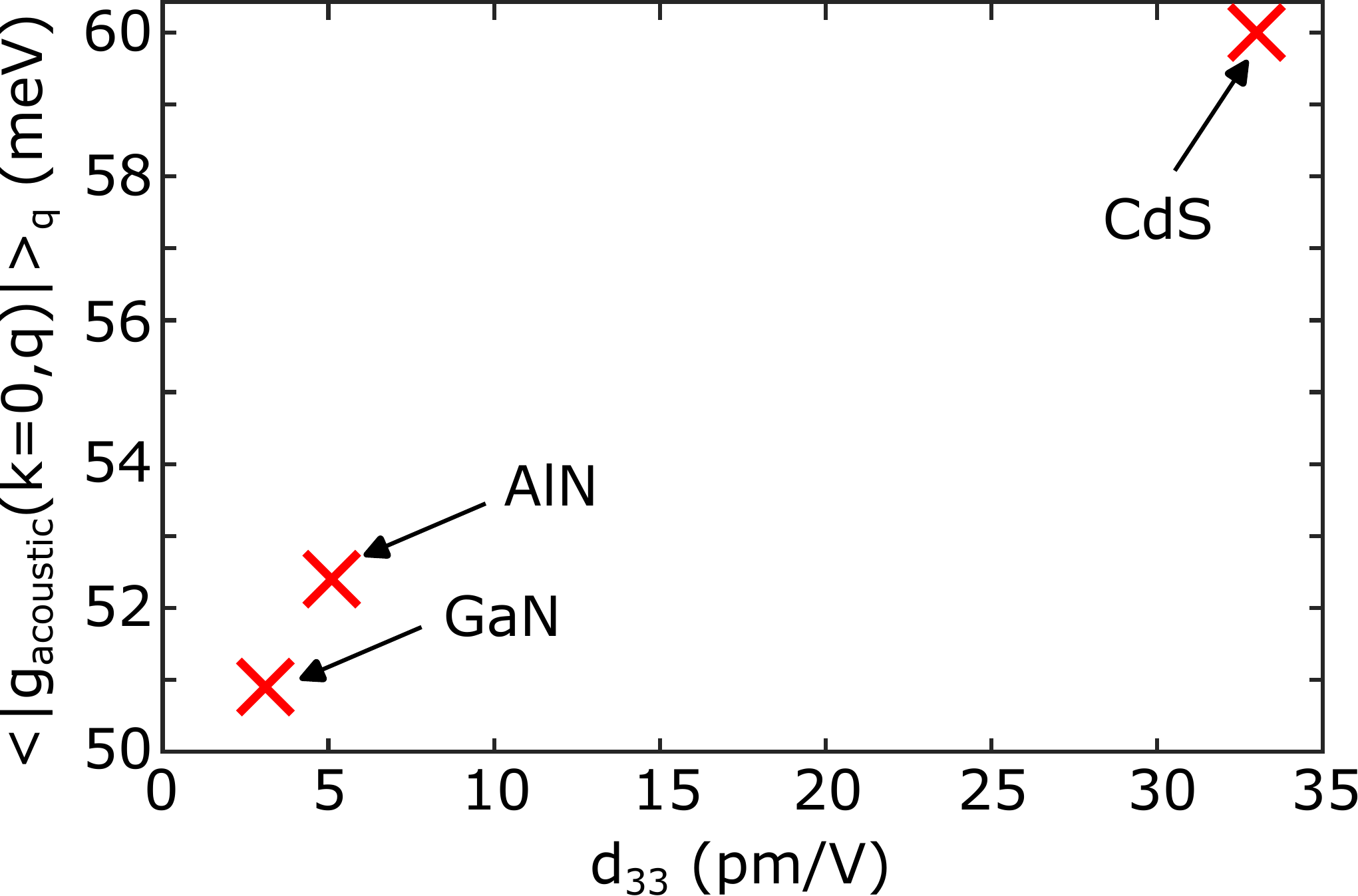}
   \caption{Average electron-phonon coupling of acoustic modes of CdS, AlN and GaN, plotted against the experimental piezoelectric coefficient (Ref.~\cite{Lueng2000} for AlN and GaN, Ref.~\cite{Wang2016} for CdS.)}
    \label{fig:piezoelectric}
\end{figure}

As seen in the previous Section\,\ref{finite_temperatures}, and specifically, in Figures \,\ref{fig:cds_deb_vs_omega} and\,\ref{fig:gan_deb_vs_omega}, acoustic phonons
can result in a substantial reduction of the exciton binding energy in CdS and
GaN. Moreover, among the systems studied in this work, AlN is the only other
case where acoustic phonons contribute to phonon screening. What 
these materials have in common
is that all three are
piezoelectric~\cite{Lueng2000,Wang2016}, enabling a large coupling of the electrons with acoustic phonons~\cite{MahanGeraldD2013MP}, with the average electron-phonon coupling of acoustic modes scaling monotonically with the
experimental piezoelectric constants, as seen in Fig.\,\ref{fig:piezoelectric}. 

As shown previously, ignoring quadrupole terms in the Wannier-Fourier interpolation 
of the electron-phonon matrix elements of acoustic
phonons can lead to an
overestimation of the magnitude of $g$ in the vicinity of 
$\Gamma$~\cite{Jhalani2020}. Therefore, it is possible that inclusion of
quadrupole corrections will lead to a modest reduction of the 
predicted screening of the exciton binding energy by acoustic phonons. 
Nevertheless, the contribution of
acoustic phonons to screening excitons will be important in 
piezoelectric materials, highlighting
the importance of going beyond the simple picture within which optical phonons
are the only ones contributing to this effect.


\section{Discussion, conclusions and outlook}
\label{discussion}

In this work we have developed an \emph{ab initio} framework for
computing the temperature-dependent phonon kernel of semiconductors and insulators, according
to Eq.\,\ref{eq:phonon_kernel}. We show
how approximations to the phonon kernel lead to the model expressions 
of Eq.\,\ref{eq:Haken_deb},\,\ref{eq:FHN_deb} and\,\ref{eq:kph_numerical}, some of which have been widely discussed
in the literature.
Compared
to utilizing these model expressions,
our \emph{ab initio} approach does not rely on any restrictive approximations, treating
the effect of all phonon modes on equal footing without any assumption about the nature of their coupling to electrons. It also allows us to
extend our results to finite temperatures in a straightforward manner, and to go beyond the Wannier-Mott model, which will be critical
in systems where excitons are not hydrogenic in nature~\cite{Biega2021}.
Additionally, the imaginary part of the phonon kernel in the exciton basis allows us to
extract information about temperature-dependent exciton dissociation processes in certain limits. Overall, having access to the full, temperature-dependent complex phonon kernel enhances our ability to predict excitonic properties while providing new physical insights, with some of the ones obtained in this work summarized
below.

Firstly, we found that for bulk CdS, phonon screening reduces the bare
BSE exciton binding energy by more than $50\%$ at room temperature, 
demonstrating that the effect of phonon screening on excitons can be
very strong and highly temperature-dependent. Our theoretical framework provides a temperature-dependent correction to the BSE kernel and therefore to the excitonic interactions. Indeed, 
experimental studies already in the 1970s and 1980s found indications of strongly temperature-dependent excitonic interactions, but were unable to quantify this effect~\cite{PhysRevB.30.1979,PhysRevB.31.947,PhysRevB.31.958}. 
These modified interactions can lead to strongly temperature-dependent exciton binding energies, as recent experimental studies have found~\cite{Miyata2015,Davies2018}, and at odds with the common assumption of a temperature-independent exciton binding energy~\cite{Jeong2000,Viswanath1998}.

Our formalism  
describes the screening of the exciton due to both processes of emission and absorption of phonons, effects that contribute equally to the reduction of the exciton binding energy
at $300$\,K, while phonon emission processes entirely dominate at low 
temperatures. Additionally, we found an important reduction of the
exciton binding energy due to screening from acoustic phonons in piezoelectric materials, which, to
the best of our knowledge, has not been previously discussed. 

Importantly, having access to the imaginary part of the phonon kernel
allows us to compute exciton dissociation rates via single-phonon emission and absorption processes entirely from first
principles. For the case of GaN where $E_B<\omega_{LO}$, the absorption of
a LO phonon is sufficient to dissociate
the exciton and generate a free electron-hole pair. Our computed exciton dissociation timescale for this system is approximately $111$\,fs at $300$\,K. While we have not found experimental exciton dissociation timescales for GaN to compare against, it is encouraging that our predicted value is reasonably close to experimental values obtained in semiconductors with comparable exciton binding energies, such as GaSe with an exciton dissociation timescale of $112$\,fs~\cite{Allerbeck2021}, and $\text{CH}_3\text{NH}_3\text{PbI}_3$ with an exciton dissociation timescale of approximately $50$\,fs~\cite{Jha2018}.

Moving forward, our first-principles approach could be extended in multiple
ways. For example, one could re-solve the BSE non-perturbatively upon correction
of the electronic BSE kernel, which might substantially change the
exciton wavefunction and computed absorption spectra for systems where
the phonon kernel in the bare exciton basis has large off-diagonal entries. Furthermore, one could
consider the interplay of phonon screening with the effects of polaronic mass enhancement and polaron
interference on excitons~\cite{Mahanti1970,Mahanti1972} by incorporating higher-order diagrams into this approach. Our approach for ensuring gauge consistency between interpolated electron-phonon matrix elements and exciton coefficients could be extended to account for the effects of phonon screening and other diagrams on finite-momentum excitons~\cite{Antonius2022}, by utilizing recent schemes for exciton Wannier functions~\cite{Haber2023}.
While in this work we have studied some representative semiconducting
materials, we hope our first-principles approach will be widely adopted
and used to study the effect of phonons on dissociating and screening
excitons in diverse materials of interest for a variety of technological
applications, such as heterostructures of two-dimensional semiconductors, quantum wells, or doped systems.

\section*{Acknowledgments}
This work was primarily supported by the
Theory FWP, which provided $GW$ and $GW$-BSE calculations and analysis of phonon effects, and the Center for Computational Study of Excited-State Phenomena in Energy Materials (C2SEPEM) as part of the Computational Materials Sciences Program, which provided advanced codes, at the Lawrence Berkeley National Laboratory, funded by the U.S. Department of Energy, Office of Science, Basic Energy Sciences, Materials Sciences and Engineering Division, under Contract No. DE-AC02-05CH11231. Z.L. and S.G.L. acknowledge support from the National Science Foundation under Grant No. OAC-2103991 in the development of interoperable software enabling the EPW and BerkeleyGW calculations with consistent gauge. Computational resources were provided by the National Energy Research Scientific Computing Center (NERSC).
M.R.F. acknowledges support from the UK Engineering and Physical Sciences Research Council (EPSRC), Grant EP/V010840/1.

\appendix

\section{Estimated exciton and polaron radii of studied systems}
\label{radii}

\begin{table*}[tb]
\centering
  \setlength{\tabcolsep}{6pt} 
\begin{tabular}{cccccc}
\hline
Material & $m_e$ (a.u.) & $m_h$ (a.u.) & $a_o$ (\AA) & $r_e$ (\AA) & $r_h$ (\AA) \\
\hline
AlN & $0.30$ & $0.70$ & $11.3$ &$10.8$ & $7.0$ \\
CdS & $0.12$ & $2.01$ & $29.4$ &$30.5$ & $7.5$ \\
GaN & $0.15$ & $1.01$ & $21.1$ &$17.3$ & $6.7$ \\
MgO & $0.34$ & $5.00$ & $6.1$ &$11.6$ & $3.0$ \\
SrTi$\text{O}_3$ & $0.39$ & $1.22$ & $10.2$ &$9.9$ & $5.6$ \\
\hline
\end{tabular}
\caption{Estimated exciton Bohr radii $a_o$, electron-/hole-polaron radii $r_{e,h}$, and electron/hole effective masses $m_{e,h}$ of the studied systems.}
\label{table:radii}
\end{table*}

Table\,\ref{table:radii} summarizes the exciton Bohr radius $a_o$ and the electron-/hole-polaron radii $r_{e,h}$ of the studied systems. The exciton bohr radius is estimated as $a_o=1/(2E_B\mu)^{1/2}$ within the Wannier-Mott model, where $E_B$ the converged BSE exciton binding energies given in Table\,\ref{table:comparison}, and $\mu$ the exciton effective mass $1/\mu=1/m_e+1/m_h$,
with $m_e$ and $m_h$ the effective mass of the electron and hole, respectively. A more quantitatively accurate calculation of the exciton radius would require using first-principles methods that
accurately capture this quantity~\cite{Sharifzadeh2013}. 
The estimated electron-/hole-polaron radii are obtained as $r_{e,h} = \frac{1}{\sqrt{2m_{e,h}\omega_{LO}}}$~\cite{Mahanti1972}, 
under the assumption of weak, Fr\"{o}hlich-like electron-phonon coupling. While recent development of first-principles methodologies now allows more accurate computation of the spatial extent of electron and hole polarons~\cite{Lafuente-Bartolome2022}, the values reported in Table\,\ref{table:radii} indicate that our studied materials are within the regime where the lattice polarization associated with the two polarons may interfere significantly, as described elsewhere~\cite{Mahanti1970,Pollmann1977}.

Moreover, Table\,\ref{table:radii} summarizes the electron and hole effective masses.
We compute the effective
masses for the top/bottom of the valence
and conduction bands respectively
using the finite difference formula $\frac{1}{m^*}=\frac{E(\delta \mathbf{k})+E(-\delta \mathbf{k})-2E(\Gamma)}{\delta \mathbf{k}^2}$, taking $\delta \mathbf{k}$ to be
equal to $0.01$ (in crystal coordinates) along each spatial direction, and we average over the three spatial directions. 
The energies $E$ are computed at the $GW$ level. 

\section{Imaginary part of the phonon kernel as an exciton dissociation rate}
\label{Imkph}

Here we outline in greater detail the arguments, using scattering theory and many-body perturbation theory techniques~\cite{MahanGeraldD2013MP,Peskin1995}, that establish the imaginary part of the diagonal elements of the phonon kernel in the unperturbed exciton basis as related to the rate of the dissociation channel for excitons into free electron-hole pairs, due to the absorption of a single phonon.

Within the $GW$-BSE formalism and accounting for phonon screening, the dynamics of a system
of excitons and phonons is described by the Hamiltonian
\begin{equation}
    \label{eq:scattering_H}
    H=(E_{c\mathbf{k}}-E_{v\mathbf{k}})\delta_{vv'}\delta_{cc'}\delta_{\mathbf{k}\mathbf{k}'}+K^{eh}_{cv\mathbf{k},c'v'\mathbf{k}'}+H_{ph}+K^{ph}_{cv\mathbf{k},c'v'\mathbf{k}'},
\end{equation}
where $H_{ph}$ is the phonon
Hamiltonian and $H_{\text{BSE}}=(E_{c\mathbf{k}}-E_{v\mathbf{k}})\delta_{vv'}\delta_{cc'}\delta_{\mathbf{k}\mathbf{k}'}+K^{eh}_{cv\mathbf{k},c'v'\mathbf{k}'}$ is the usual bare BSE Hamiltonian. 

At temperature $T$ the initial (bare exciton) state of Fig.\,\ref{fig:dissociation_schematic} is a particular eigenstate of the Hamiltonian $H_i=H_{\text{BSE}}+H_{ph}:$
\begin{equation}
    \label{eq:Hi}
    H_i\ket{S,N_B+1}=E_i^S\ket{S,N_B+1},
\end{equation}
with energy $E_i^S=\Omega_S+(N_B+1)\omega_{LO}$. We consider here the example of an LO phonon with occupation $N_B$ at temperature $T$, but the same holds for any phonon mode. The
absorption of a single phonon can lead to a final (free electron-hole) state, which also can be expressed as an eigenstate of the distinct but related Hamiltonian $H_f=(E_{c\mathbf{k}}-E_{v\mathbf{k}})\delta_{vv'}\delta_{cc'}\delta_{\mathbf{k}\mathbf{k}'}+H_{ph}$, where
\begin{equation}
    \label{eq:Hf}
    H_f\ket{(c\mathbf{k},v\mathbf{k}'),N_B}=E_f\ket{(c\mathbf{k},v\mathbf{k}'),N_B},
\end{equation}
with $E_f = E_{c\mathbf{k}}-E_{v\mathbf{k}'}+N_B\omega_{LO}$. 
The initial and final states are described as eigenstates of 
different Hamiltonians, apparently complicating 
a straightforward
interpretation of the scattering process, as one needs to ensure orthogonality between these
wavefunctions~\cite{Perea-Causin2021}. However, this complication is resolved using the arguments of the theory of rearrangement collisions~\cite{Lippmann1956,Sunakawa1960,Day1961}. We can define a renormalized
final free electron-hole state $\ket{\chi_f}$, which satisfies orthogonality to the initial state $\ket{S,N_B+1}$, as~\cite{scattering}
\begin{align}
    \label{eq:mod_state}
    \ket{\chi_f}=\ket{(c\mathbf{k},v\mathbf{k}'),N_B}\\ \nonumber+(E_f-H-i\eta)^{-1}K^{eh}\ket{(c\mathbf{k},v\mathbf{k}'),N_B}.
\end{align}
State $\ket{\chi_f}$ may be used to define
a generalized $\mathcal{S}$-matrix for the scattering process, which to first
order in the electron-phonon interaction and within the Born approximation is written as
\begin{equation}
    \mathcal{S}_{gen}^{Born}=2\pi i\delta(E_f-E_i)\bra{(c\mathbf{k},v\mathbf{k}'),N_B}K^{ph}\ket{S,N_B+1}.
\end{equation}
Employing the optical theorem for this  $\mathcal{S}$-matrix in standard fashion, $\text{Im}[K^{ph}_{SS}(\Omega_S,T)]$ can be interpreted as the rate of this exciton dissociation process, namely
\begin{equation}
    2|\text{Im}[K^{ph}_{SS}(\Omega_S,T)]|\approx\tau^{-1}_S(T).
\end{equation}

\section{Convergence of the first-principles phonon kernel}
\label{appendix_convergence}

\begin{figure}[tb]
    \centering
    \includegraphics[width=0.9\linewidth]{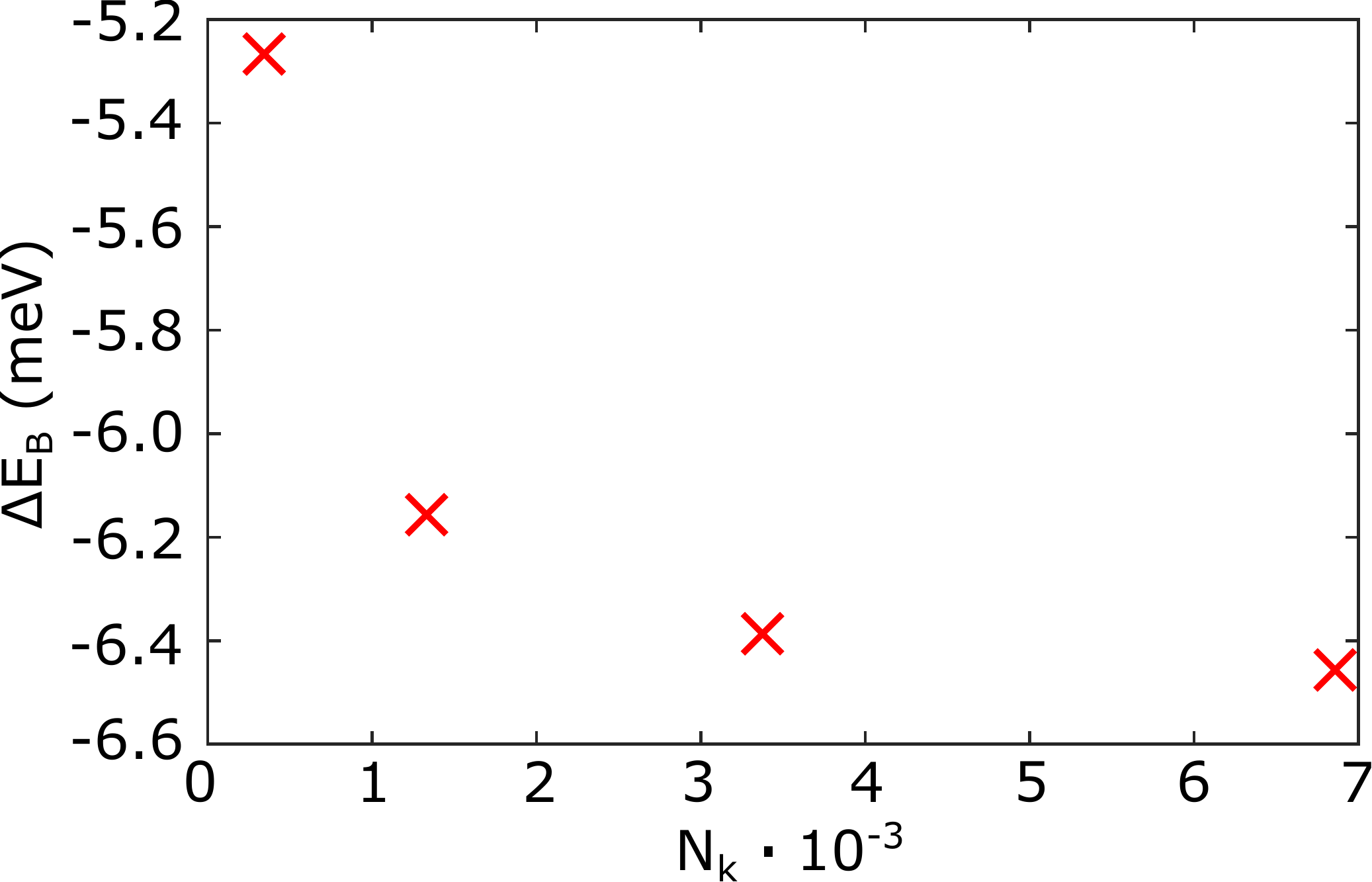}
    \caption{Convergence of the exciton binding energy shift of CdS due to phonon screening at $0$\,K, with respect to the number of k-points in a patch centered around the $\Gamma$-point of a $100\times100\times100$ grid.}
    \label{fig:cds_deb_convergence}
\end{figure}

\begin{figure}[tb]
    \centering
    \includegraphics[width=0.9\linewidth]{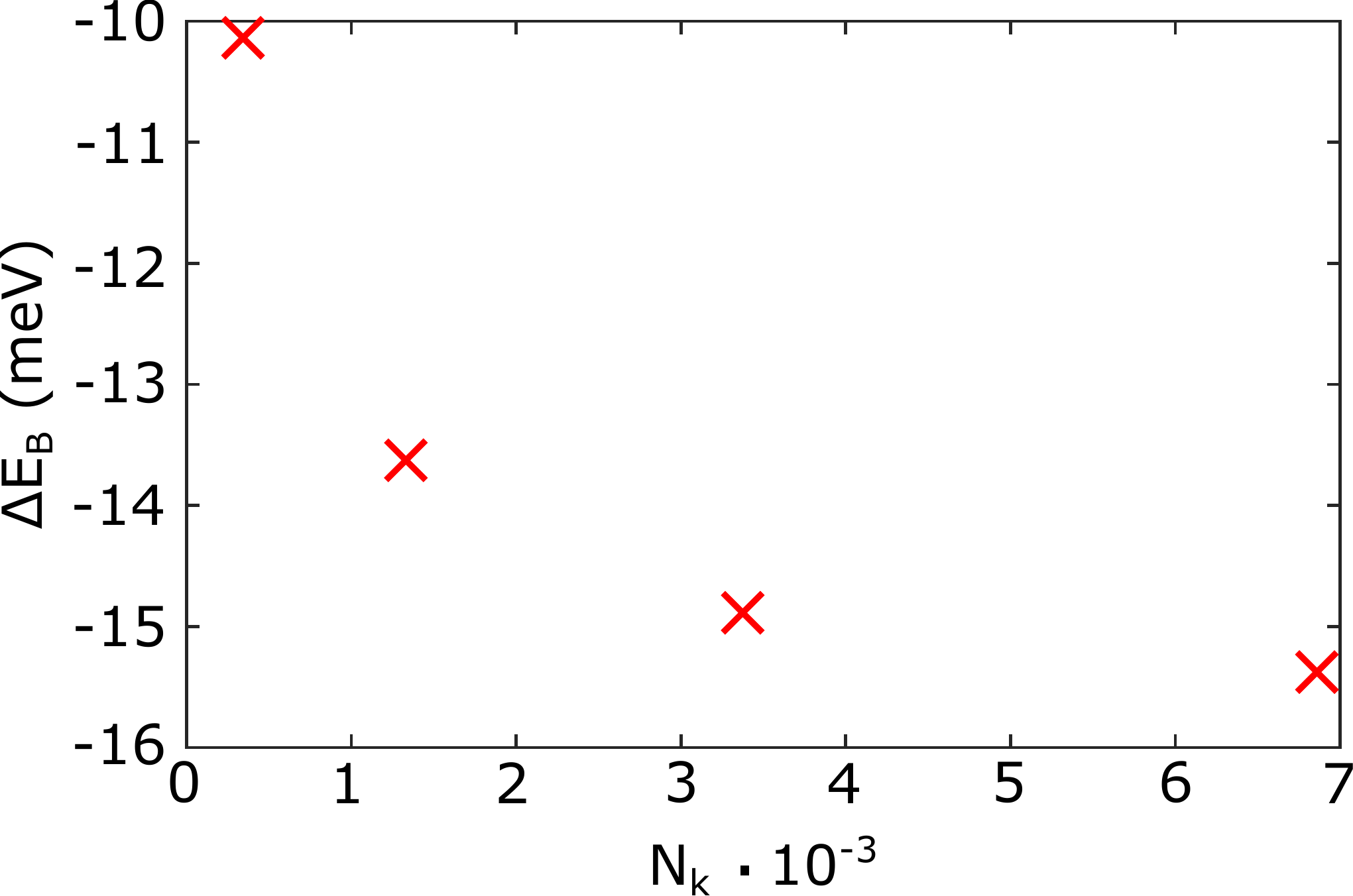}
    \caption{Convergence of the exciton binding energy shift of GaN due to phonon screening at $0$\,K, with respect to the number of k-points in a patch centered around the $\Gamma$-point of a $100\times100\times100$ grid.}
    \label{fig:gan_deb_convergence}
\end{figure}

Figures\,\ref{fig:cds_deb_convergence} and\,\ref{fig:gan_deb_convergence}
demonstrate the convergence of
the real part of $K^{ph}$ in CdS and GaN respectively, \emph{i.e.} the correction $\Delta E_B$
to their exciton binding energy from phonons. By employing a patch taken from a $100\times 100 \times 100$ regular grid, we are able to demonstrate convergence within less than $1$\,meV at $0$\,K, for a patch cutoff of  $0.09$ (in crystal coordinates) around $\Gamma$, corresponding to $6,859$ $\mathbf{k}/\mathbf{q}$-points.

\begin{figure}[tb]
    \centering
    \includegraphics[width=0.9\linewidth]{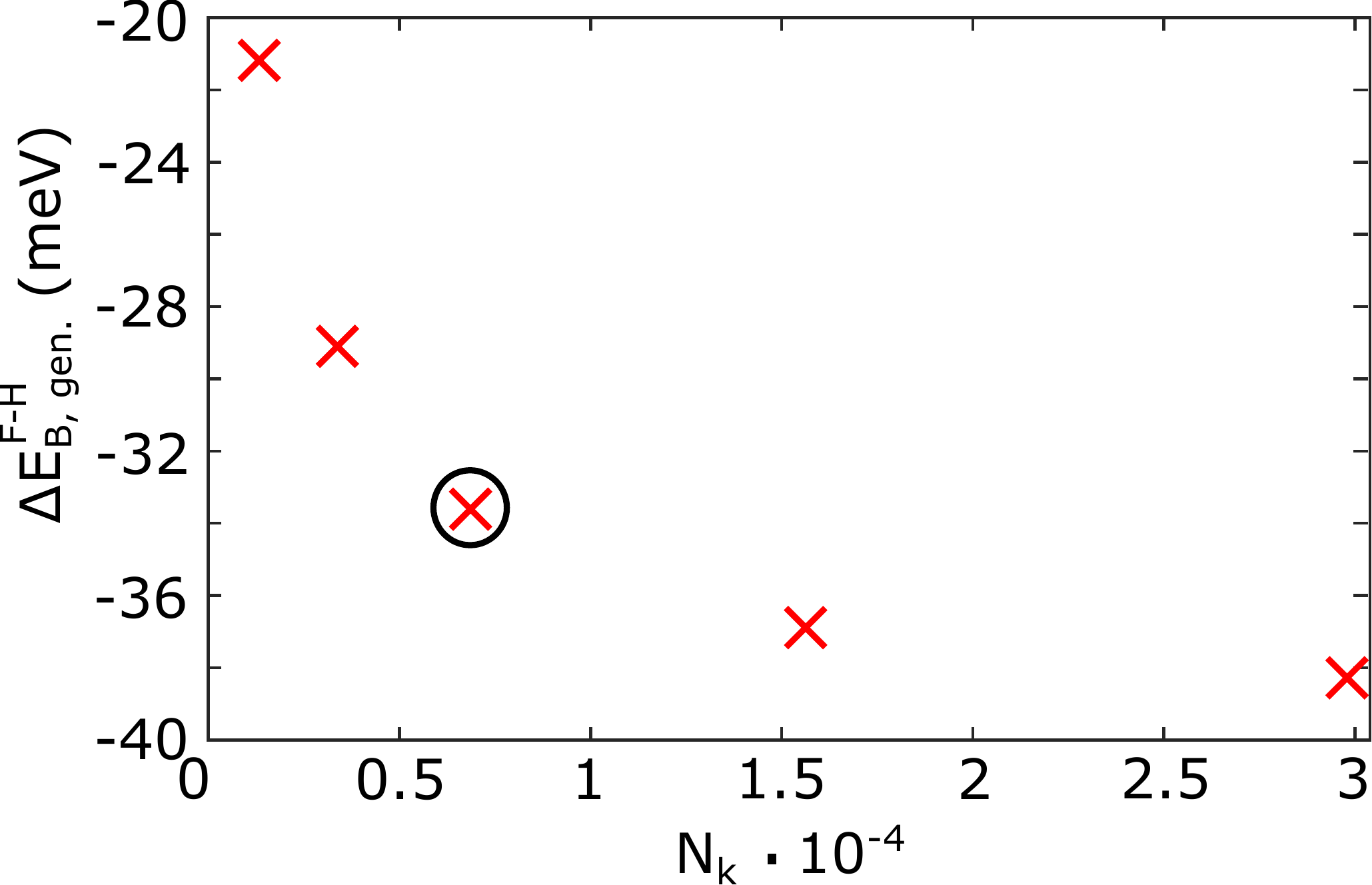}
    \caption{Convergence of the exciton binding energy shift of $\text{SrTiO}_3$ due to phonon screening from the LO-1 mode at $0$\,K, with respect to the number of k-points in a patch centered around the $\Gamma$-point of a $100\times100\times100$ grid, as computed through numerical integration of the approximate Eq.\,\ref{eq:kph_numerical} employing the generalized Fr\"{o}lich vertex of Eq.\,\ref{eq:long_range_g}. The point in the black circle corresponds to the maximum grid size which was accessible through our \emph{ab initio} workflow, and which is extrapolated to the $N_k\rightarrow \infty$ limit as discussed in the text.}
    \label{fig:STO_deb_convergence}
\end{figure}

For $\text{SrTiO}_3$ convergence is more challenging to achieve. Fig.\,\ref{fig:STO_deb_convergence} demonstrates the values of $\Delta E_{B,\text{gen.}}^{\text{F-H}}$ contributed by the LO-1 phonon, as predicted by Eq.\,\ref{eq:kph_numerical} when using the generalized Fr\"{o}hlich vertex of Eq.\,\ref{eq:long_range_g}. In order to fully converge $\Delta E_B$ we have to compute the sum of Eq.\,\ref{eq:kph_numerical} on a grid of $29,791$ $\mathbf{k}/\mathbf{q}$-points (corresponding to a patch with a cutoff of $0.15$ around $\Gamma$ in crystal coordinates, drawn from a $100\times 100\times 100$ regular grid), which is possible to do within this numerical
integration employing the hydrogenic and Fr\"{o}hlich formulas. However, convergence at the same level
is very challenging to achieve from first principles. When employing our \emph{ab initio} workflow, we were able to compute the phonon kernel on a $\Gamma$-centered patch with a cutoff of $0.09$ (crystal coordinates), which includes $6,859$ $\mathbf{k}/\mathbf{q}$-points, obtaining $\Delta E_B$ values of $-31.1$\,meV and $-7$\,meV for the LO-1 and LO-2 phonons respectively. For the same grid size, the numerical integration of Eq.\,\ref{eq:kph_numerical} gives corrections of $-33.6$\,meV and $-7$\,meV
for these two phonons respectively. Therefore, for the LO-2 phonon the two
levels of theory are in perfect agreement at $0$\,K, whereas for the LO-1
phonon, the first-principles workflow gives a correction which is equal
to $92.7\%$ of the $\Delta E_{B,\text{gen.}}^{\text{F-H}}$ value of Fig.\,\ref{fig:STO_deb_convergence} on the same grid size (circled in black). Consequently, in order to estimate
the \emph{ab initio} values at the $N_k\rightarrow \infty$ limit, for
the LO-1 phonon contribution to $\Delta 
E_{B}^{ab\hspace{0.1cm} initio}$, we take $92.7\%$ of the
converged value of $\Delta E_{B,\text{gen.}}^{\text{F-H}}$ as estimated from Eq.\,\ref{eq:kph_numerical} ($-38$\,meV), which is equal to $-35$\,meV (Table\,\ref{table:STO_updated_model}). For the LO-2 phonon the \emph{ab initio} and the generalized Fr\"{o}hlich-hydrogenic results on the patch of $6,859$ $\mathbf{k}/\mathbf{q}$-points are identical, we therefore simply extrapolate the first-principles contribution to $\Delta E_B$ from this phonon from the slightly under-converged value of $-7$\,meV, to the $N_k\rightarrow \infty$ limit of $-8$\,meV. This same method for extrapolating to $N_k\rightarrow \infty$ is used for the phonon screening of the MgO exciton by the LO phonon of this material. 

\section{Effect of thermal expansion}
\label{thermal_expansion}

\begin{table*}[tb]
\centering
  \setlength{\tabcolsep}{6pt} 
\begin{tabular}{cccc}
\hline
Material & Thermal Expansion Coefficient ($\text{K}^{-1}$) & Reference & $\Delta a/a(\Delta T=300\,\text{K})$ (\%)\\
\hline
AlN & $6.9\cdot 10^{-6}$ & \cite{Kroncke2008} & $0.2$\\
CdS & $4.5\cdot 10^{-6}$ & \cite{Strauch2012} & $0.14$\\
GaN & $6\cdot 10^{-6}$ & \cite{Roder2005} & $0.18$\\
MgO & $3.2\cdot 10^{-5}$ & \cite{Corsepius2007} & $1$\\
SrTi$\text{O}_3$ & $3\cdot 10^{-5}$ & \cite{Boudali2009} & $0.9$ \\
\hline
\end{tabular}
\caption{Thermal expansion coefficients of the studied systems, as found in the literature, and percentage of change to the $a$ lattice constant assuming linear thermal expansion over $300$\,K.}
\label{table:te_coefficients}
\end{table*}

In Table\,\ref{table:te_coefficients} we report the
coefficients of linear thermal expansion of the systems we study here, as found in the literature. 
For $\text{SrTiO}_3$ we take a representative value
for room temperature as reported for the zero external pressure case in Ref.~\cite{Boudali2009}. Assuming linear thermal expansion over the temperature range $0-300$\,K, we estimate the percentage of change $\Delta a/a$ to the lattice constant over these temperatures for each material. We see in Table\,\ref{table:te_coefficients} that for AlN, CdS, and GaN, the lattice constants change by at most $0.2$\%, we therefore neglect this effect for the range of temperatures we are concerned with here. 
For MgO and $\text{SrTiO}_3$ there is a more significant change of approximately $1$\% to the lattice constant. For these two materials we have repeated the calculations of the bare and phonon-screened exciton binding energy, in order to evaluate the importance of thermal expansion. 

Starting with MgO, expanding its lattice constant by $1$\% decreases the bare exciton binding energy of
this system to $305$\,meV, compared to the $327$\,meV value of Table\,\ref{table:comparison}. The LO phonon frequency also decreases from $84$\,meV to $81$\,meV. However, the  \emph{ab initio} phonon screening of the exciton remains almost unchanged, with the effect reduced by a mere $0.5$\,meV compared to the value of Table\,\ref{table:comparison}. 

For $\text{SrTiO}_3$, upon expansion of the lattice by $1$\%, the bare exciton binding energy
reduces from $122$\,meV to $107$\,meV, while the LO-1 phonon also reduces its energy from $98$\,meV to $94$\,meV. While we were not able to obtain fully first-principles values for the phonon screening of the exciton of $\text{SrTiO}_3$ at this modified configuration due to the emergence of imaginary phonons, taking the $\mathbf{q}\rightarrow \mathbf{0}$ limit and using the model of Eq.\,\ref{eq:FHN_deb}, we obtain $\Delta E^{\mathbf{q}\rightarrow \mathbf{0}}_B=-61$\,meV, compared to $\Delta E^{\mathbf{q}\rightarrow \mathbf{0}}_B=-65$\,meV when not accounting for thermal expansion, which suggests a $6\%$ decrease in the effect of phonon screening on the exciton binding energy.

Overall, we find that the effect of thermal
expansion can lead to a small decrease of the screening of the exciton from phonons. This change in phonon screening is generally modest due to the fact that \emph{both} $\omega_{LO}$ and $E_B$ are reduced upon
expansion of the lattice, making the change in their
ratio less significant. 
The reduction of the exciton binding energy due
to thermal expansion, combined with the small effect of thermal expansion on phonon screening, will result to an overall larger decrease
of exciton binding energies when thermal expansion
and phonon screening are accounted for concurrently, and lead to better agreement with experimental values. Nevertheless, the temperature-dependent reduction of exciton binding energies due to phonon screening remains the dominant effect, as seen here even for MgO and $\text{SrTiO}_3$, the systems with the greatest effect of thermal expansion among the studied ones.

\section{Comparison of Fr\"{o}hlich-hydrogenic approximations to $\Delta E_B$ to fully \emph{ab initio} values in $\text{SrTiO}_3$}
\label{STO_comparison}

\begin{figure}[tb]
    \centering
    \includegraphics[width=0.9\linewidth]{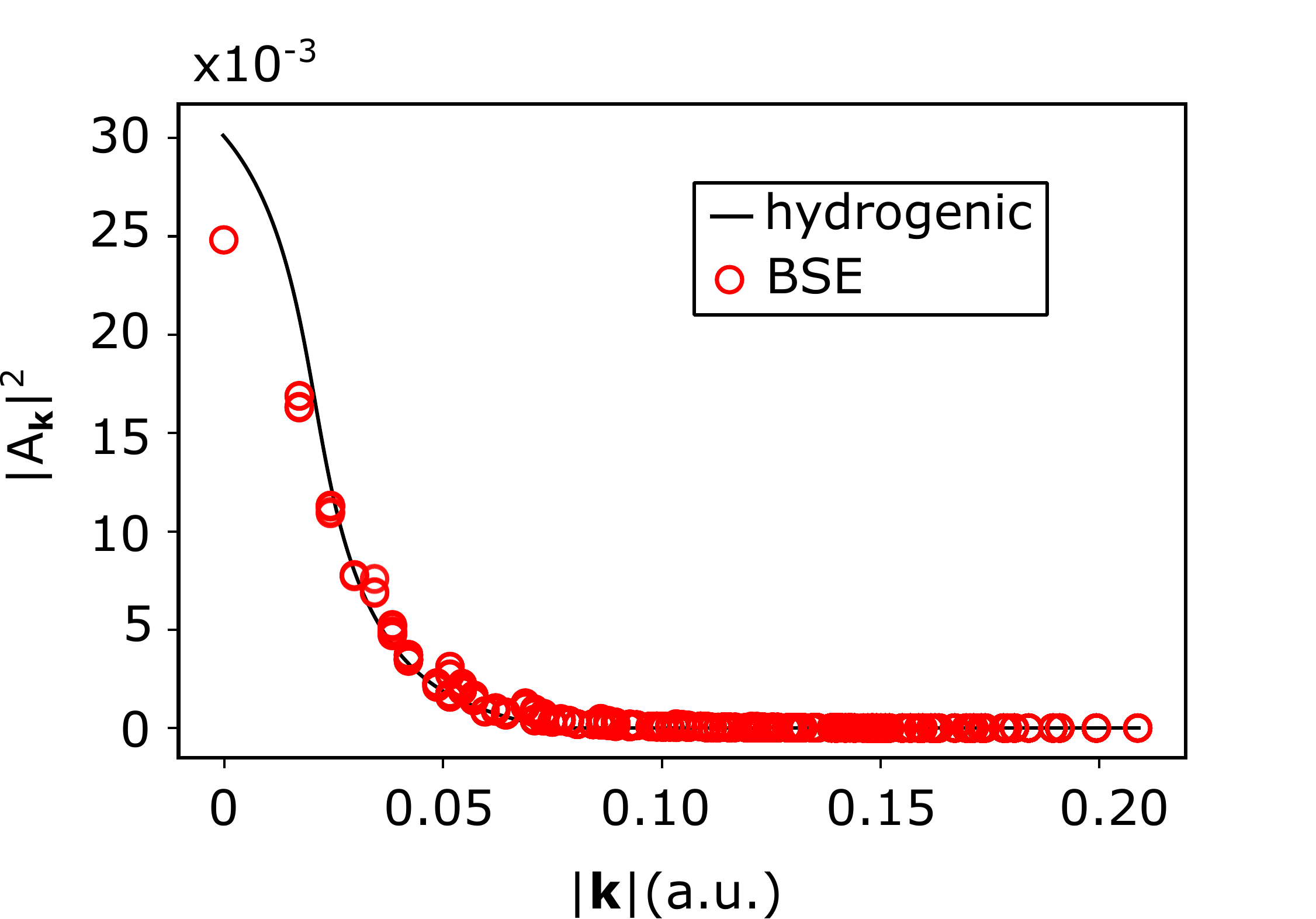}
   \caption{Comparison of the exciton wavefunction in reciprocal space computed within \emph{ab initio} $GW$-BSE and the hydrogenic model for $\text{SrTiO}_3$.}
    \label{fig:STO_hydrogenic_comparison}
\end{figure}

\begin{figure}[tb]
    \centering
    \includegraphics[width=0.9\linewidth]{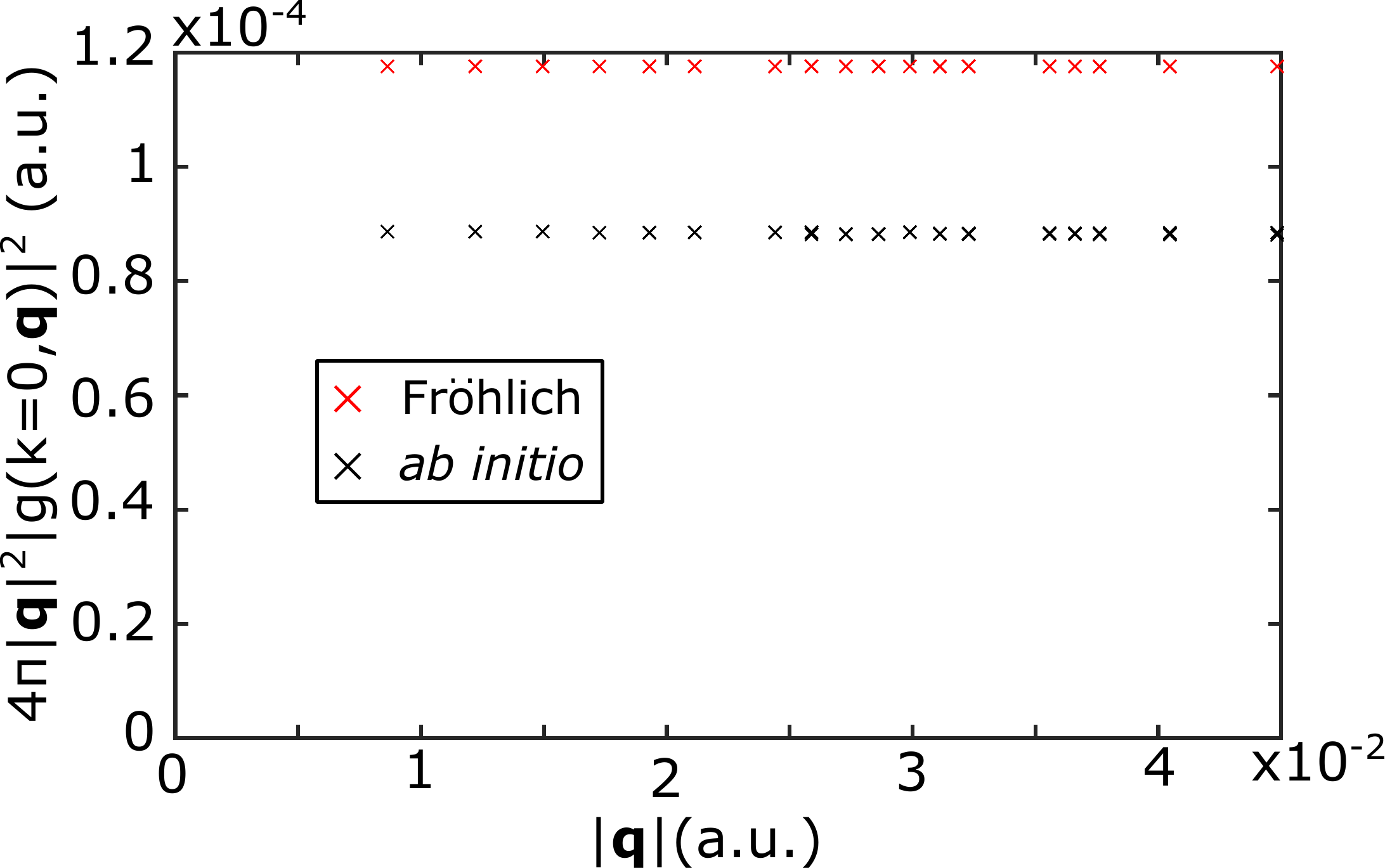}
   \caption{Comparison of the \emph{ab initio} and Fr\"{o}hlich electron-phonon coupling $g$ of SrTi$\text{O}_3$.}
    \label{fig:STO_g_comparison}
\end{figure}

\begin{figure}[tb]
    \centering
    \includegraphics[width=0.9\linewidth]{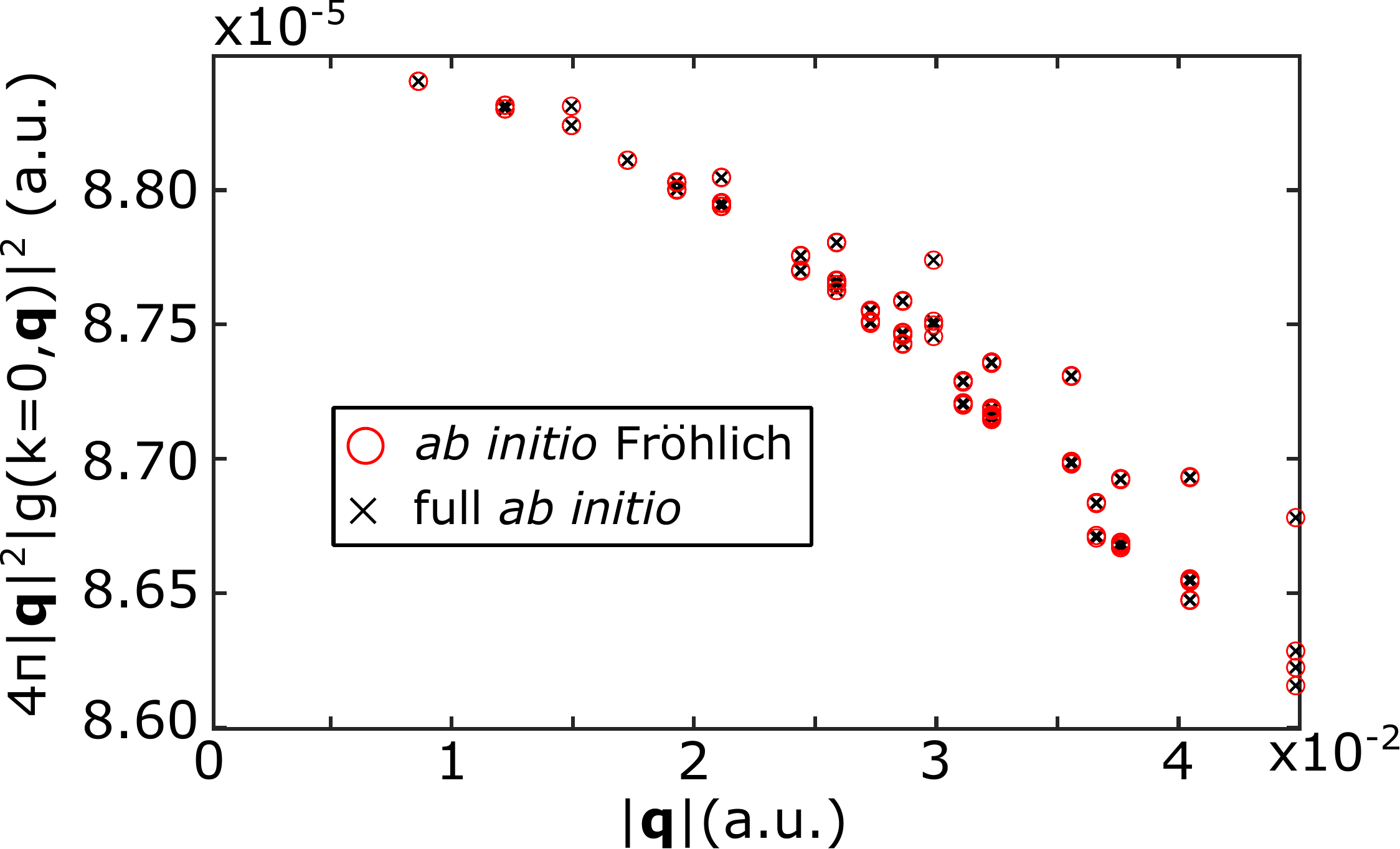}
   \caption{Comparison of the \emph{ab initio} long-range Fr\"{o}hlich coupling $g$ of Ref.~\cite{Verdi2015} and the full \emph{ab initio} electron-phonon coupling of SrTi$\text{O}_3$ for the LO-1 phonon.}
    \label{fig:STO_g_long_range_comparison}
\end{figure}

\begin{table}[tb]
\centering
  \setlength{\tabcolsep}{8pt} 
\begin{tabular}{cc}
\hline
$\omega_{m,\mathbf{q}=\mathbf{0}}$ (meV) &  $\Delta \epsilon_m$  \\
\hline
$20$ & $391.08$ \\
$57$  & $9.68$ \\
$98$ & $2.18$ \\
\hline
\end{tabular}
\caption{Phonons with finite contributions to the static dielectric constant $\epsilon_0$ of $\text{SrTiO}_3$.}
\label{table:STO_eps_decomposition}
\end{table}

The model of Eq.\,\ref{eq:kph_numerical} only includes the effect
of the highest frequency LO phonon, of
$98$\,meV. One could therefore expect that comparing the value of $\Delta E_B$ obtained through the numerical integration of Eq.\,\ref{eq:kph_numerical} ($-51$\,meV) to the \emph{ab initio} value which is just
due to the highest frequency LO phonon ($-36$\,meV at $0$\,K) might improve agreement, but in fact it only makes the agreement between the
two worse. The derivation of Eq.\,\ref{eq:kph_numerical} from the full expression of
Eq.\,\ref{eq:phonon_kernel}, relies on two approximations. Specifically,
excitons are assumed to be hydrogenic as described within the Wannier-Mott picture, and the electron-phonon interaction is considered to be
entirely dominated by the long-range Fr\"{o}hlich coupling of electrons
to LO phonons. Therefore, at least one of these two approximation has
to be violated for this system in order to explain the discrepancy
between the Fr\"{o}hlich-hydrogenic $\Delta E_B^{\text{F-H}}$ and first-principles $\Delta 
E_{B}^{ab\hspace{0.1cm} initio}$ values of Table\,\ref{table:comparison}.

We first examine the validity of the hydrogenic model in $\text{SrTiO}_3$. In Fig.\,\ref{fig:STO_hydrogenic_comparison} we compare
the exciton wavefunction as computed within the BSE to the hydrogenic
model described by Eq.\,\ref{eq:hydrogenic}, on a $\Gamma$-centered patch with a cutoff of $0.15$ (in crystal coordinates), which is necessary to satisfy the normalization of the wavefunction within $1\%$. We see that the two are in good agreement, with some small deviations close to the zone center. 

We now move to examine the agreement between the first-principles
electron-phonon matrix elements $g$, and those predicted by the Fr\"{o}hlich model $g_{Fr}$, for the LO-1 phonon, in Fig.\,\ref{fig:STO_g_comparison}, for a small region close to $\Gamma$. It is immediately apparent that
the Fr\"{o}hlich model consistently overestimates the first-principles
values for the electron-phonon matrix elements, with the ratio $\frac{g^2_{Fr}}{g^2}$ equal to approximately $4/3$. Since the
correction to the exciton binding energy is proportional to $|g_{\mathbf{q}\nu}|^2$ it is reasonable to expect a similar
overestimation, compared to the \emph{ab initio} value. Indeed, if we naively multiply the first-principles value of $-36$\,meV for the LO-1 phonon by $4/3$ we get $\Delta E_B=-48$\,meV, in very close agreement to the value of $-51$\,meV predicted by the numerical integration of Eq.\,\ref{eq:kph_numerical}.
Therefore most of the discrepancy can be attributed to the deviations
of the Fr\"{o}hlich vertex from its \emph{ab initio} value, with the
remaining difference likely due to the small differences between
the hydrogenic and BSE exciton wavefunctions, see Fig.\,\ref{fig:STO_hydrogenic_comparison}. 

We now examine the reason behind this observed failure
of the Fr\"{o}hlich model. Unlike other systems studied here and presented in Table\,\ref{table:comparison}, $\text{SrTiO}_3$ is the only
material with more than a single LO phonon contributing to phonon screening, as seen in Fig.\,\ref{fig:STO_deb_vs_omega}, as well as the
only system among the studied ones where more than a single phonon mode $m$
contributes to the low-frequency dielectric constant $\epsilon_0$.
Following Ref.~\cite{Fennie2003} 
we decompose $\epsilon_0$ as $\epsilon_0=\epsilon_{\infty}+\sum_m \Delta \epsilon_m$, and the phonon modes $m$ with finite $\Delta \epsilon_m$ contribution are given in Table\,\ref{table:STO_eps_decomposition}, as computed from DFPT.
The largest contribution to $\epsilon_0$ comes from the 
soft polar mode with a frequency of $20$\,meV at $\Gamma$. This phonon
does not contribute to the phonon screening of the exciton binding
energy, due to its weak electron-phonon coupling around $\Gamma$~\cite{Zhou2018}, where
the exciton coefficients are finite, and also because of its significantly lower
energy compared to the exciton binding energy of $122$\,meV. 

This large contribution to $\epsilon_0$ by a phonon that ultimately does
not contribute to phonon screening suggests that using $\epsilon_0$ to
estimate the Fr\"{o}hlich coupling of the LO-1 and LO-2 modes will 
lead to an overestimation of the vertex $g$ compared to a full first-principles calculation, as indeed we found in Fig.\,\ref{fig:STO_g_comparison}. The standard Fr\"{o}hlich vertex was generalized in Ref.~\cite{Verdi2015} to describe long-range
electron-phonon interactions in anisotropic materials and within a mode-resolved picture for each phonon, as given in Eq.\,\ref{eq:long_range_g}.
This so-called \emph{ab initio} Fr\"{o}hlich vertex,
perfectly reproduces the full first-principles values of the electron-
phonon matrix elements close to $\mathbf{q}=\mathbf{0}$, as
shown in Fig.\,\ref{fig:STO_g_long_range_comparison} for the LO-1
phonon, and the same being true for the LO-2 phonon. This suggests
that using Eq.\,\ref{eq:long_range_g} to describe the electron-phonon
matrix element in Eq.\,\ref{eq:kph_numerical} will restore good 
agreement between the first-principles $\Delta 
E_{B}^{ab\hspace{0.1cm} initio}$ and numerically-integrated value 
of $\Delta E_{B,\text{gen.}}^{\text{F-H}}$, with the latter now based on the hydrogenic model for 
excitons and the \emph{generalized} Fr\"{o}hlich model. Indeed in 
Table\,\ref{table:STO_updated_model} we show that by applying these two 
levels of theory to the phonon modes LO-1 and LO-2, we recover close-to-
perfect agreement between the results of Eq.\,\ref{eq:kph_numerical} and
taking the real part of Eq.\,\ref{eq:phonon_kernel}. Any remaining
discrepancy is attributed to the differences between the hydrogenic
model and BSE in terms of describing the exciton wavefunction, see Fig.\,\ref{fig:STO_hydrogenic_comparison}.

\textbf{}

\end{document}